\documentclass[aps,prd,showpacs,twocolumn]{revtex4}
\usepackage{graphics}
\usepackage{amssymb}
\usepackage{amsmath}
\usepackage{epsfig}
\usepackage{epsf}
\usepackage{slashed}
\usepackage[usenames]{color}
\begin{document}

\title{Cooling curves for neutron stars with hadronic matter and quark matter}

\author{Shaoyu Yin$^1$\footnote{s.yin@uu.nl}, J.J.R.M. van Heugten$^1$,
Jeroen Diederix$^1$, Maarten Kater$^1$, Jacco Vink$^2$, and H.T.C. Stoof$^1$}\affiliation{\small
1. Institute for Theoretical Physics, Utrecht University, Leuvenlaan 4, 3584 CE Utrecht, The Netherlands\\
\small 2. Astronomical Institute, Utrecht University, P.O. Box 80000, 3508 TA Utrecht, The Netherlands}

\begin{abstract}
The thermal evolution of isothermal neutron stars is studied with
matter both in the hadronic phase as well as in the mixed phase of
hadronic matter and strange quark matter. In our models, the
dominant early-stage cooling process is neutrino emission via the
direct Urca process. As a consequence, the cooling curves fall too
fast compared to observations. However, when superfluidity is
included, the cooling of the neutron stars is significantly slowed
down. Furthermore, we find that the cooling curves are not very
sensitive to the precise details of the mixing between the
hadronic phase and the quark phase and also of the pairing that
leads to superfluidity.
\end{abstract}


\maketitle

\section{Introduction}

Neutron stars are natural laboratories for physics under extreme
conditions with their extremely high densities, powerful energy
emission, large magnetic fields, and millisecond rotation periods.
At the densities near the surface of such a star, atoms break
apart into nuclei and electrons. At higher densities, the
electrons neutralize with the protons in the nuclei to form
neutrons. These stars thus consist of a large fraction of neutrons
and are supported from gravitational collapse by the neutron
degeneracy pressure, from which the neutron star derives its name.

However, at high densities the existence of more exotic particles
is expected. These particles are generated by processes which
produce strangeness, such as
\begin{equation}\label{lambda}
n+n\rightarrow n+\Lambda^0+K^0,
\end{equation}
where $n$ is the neutron, $\Lambda^0$ the Lambda hyperon, and
$K^0$ the strange meson. The strange meson can decay via various
weak processes and the final products, usually photons and
neutrinos, leak out of the star. Therefore, the reverse process is
reduced and some net amount of strangeness survives in the dense
core of the star \cite{Glendenning:1985}. It is generally believed
that these strangeness carrying particles, called hyperons, can
exist in the center of neutron stars. They coexist with the
nucleons as well as the leptons $e^-$ and $\mu^-$. The
interactions between them are dominated by the complicated nuclear
force whose carriers are the mesons. We refer to such a system as
the hadronic phase of matter. The system with only nucleons and
leptons, i.e., without hyperons, is referred to as the nuclear
phase of matter.

\begin{figure}
\resizebox{1\linewidth}{!}{\includegraphics*{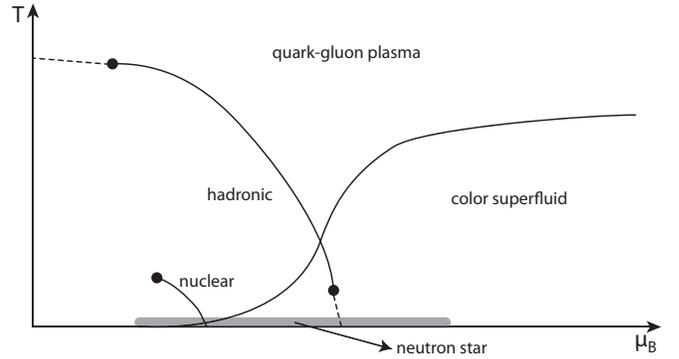}}
\caption{The schematic QCD phase diagram as a function of temperature $T$ and baryon chemical potential $\mu_B$. The solid lines denote first- and second-order phase transitions, whereas the dashed lines denote smooth crossovers. According to the typical temperature of a proto-neutron star of $T \sim 10^{10}$ K $\sim 1$
MeV, the state of matter in the neutron star should be close to
the bottom line, as indicated by the grey thick line.}\label{QCD}
\end{figure}

At even higher densities, as a consequence of asymptotic freedom,
quarks become deconfined from the hadrons. Therefore, strange
quark matter, which consists of $u$, $d$ and $s$ quarks, may also
exist in the neutron star core. There may also be a
phase-separated mixture of strange quark matter and hadronic matter in a
certain range of densities \cite{Glendenning:1992}, which we call
the mixed phase. At present, there is still a lot unknown about
the deconfinement phase transition of quarks. The contemporary
knowledge on matter at high densities and temperatures is shown by
the schematic QCD phase diagram in Fig.~\ref{QCD}. Since the
baryon chemical potential $\mu_B$ is a monotonously increasing
function from the surface to the center of the neutron star, the
$\mu_B$ axis can be mapped to the stellar radius and the different
phases in the neutron star are explicitly indicated.

Because of the limited knowledge on the state of matter at high
densities and the complexity of the interactions between the
particles, many effective models for matter inside neutron stars
have been constructed. In general, these models supply an equation
of state, which determines the particle composition of the neutron
star. The observation of neutron stars, in turn, constrain these
models. For example, if an equation of state is too soft it is
incapable of supporting a very large stellar mass, such that some
models become disfavored whenever the data on the heaviest neutron
star is updated. The stellar mass is not the only constraint on
the models. The cooling of neutron stars can be studied from their
luminosity as a function of time. A proto-neutron star is born
with a typical temperature larger than $10^{10}\textrm{ K}$, after
which it mainly loses its energy by two processes, namely by
neutrino emission everywhere inside the star and by photon
radiation at the surface. In the early stages of the thermal
evolution of the star, neutrino emission is the dominant cooling
effect after which photon radiation ultimately takes over
\cite{Yakovlev:2001}. Since neutrino emission occurs everywhere in
the neutron star, it provides a probe for studying the state of
matter inside the star.

The most efficient neutrino-emission process is called the direct
Urca (DUrca) process
\begin{equation}\label{DUrca}
n\rightarrow p+e^-+\bar{\nu}_e,\quad p+e^-\rightarrow n+\nu_e.
\end{equation}
This process is only possible if the proton fraction is more than
a certain threshold in order for it to satisfy energy and momentum
conservation \cite{Lattimer:1991}. Historically, the proton
abundance in neutron stars was underestimated. In that scenario,
it is reasonable to take into account also the modified Urca
(MUrca) process, where a bystander helps the momentum
conservation, for example
\begin{align}\label{MUrca}
n+n\rightarrow p+n+e^-+\bar{\nu}_e,&\ p+n+e^-\rightarrow n+n+\nu_e,\nonumber\\
n+p\rightarrow p+p+e^-+\bar{\nu}_e,&\ p+p+e^-\rightarrow n+p+\nu_e.
\end{align}
Some other processes can have a neutrino emissivity which is
smaller or comparable to the modified Urca process, such as
neutrino bremsstrahlung and plasmon decay \cite{Yakovlev:2001}.
However, all of them are negligible whenever the direct Urca
channel is open. In the hadronic phase, the direct Urca processes
are quite rich and can be summarized as
\begin{equation}\label{DUrcabaryon}
b_1\rightarrow b_2+l^-+\bar{\nu}_l,\quad b_2+l^-\rightarrow b_1+\nu_l,
\end{equation}
where $b_1$ and $b_2$ denote two different baryons and $l^-$
represents one of the leptons, $e^-$ or $\mu^-$.

For strange quark matter, the direct Urca processes are
\begin{align}\label{DUrcaquark}
d\rightarrow u+l^-+\bar{\nu}_l,\quad u+l^-\rightarrow d+\nu_l,\nonumber\\
s\rightarrow u+l^-+\bar{\nu}_l,\quad u+l^-\rightarrow s+\nu_l,
\end{align}
which are simply the direct Urca processes of the baryons at the
quark level. Inside the neutron star, there is no threshold for
the direct Urca process in quark matter, since the Fermi momenta
of the two quarks and lepton can always satisfy momentum conservation. As
mentioned, the neutrino emissivity depends strongly on the type of
matter contributing to the process, such that the thermal
evolution of a star directly probes its composition. The main goal
of this paper is to study the cooling of neutron stars with
different types of equations of state and try to constrain them by
comparing with observational data.

A large amount of research on the cooling process of neutron stars
has been carried out in the last several decades, with a strong
focus on the nuclear phase of matter. To the best of our
knowledge, a unifying model covering the low-density nuclear
phase to the deconfined quark phase, consistent with the QCD phase
diagram, has been less thoroughly explored. The strange quark star
composed of strange matter, either with or without the nuclear
crust, has been considered in some papers
\cite{Schaab:1996,Page:2002}. However, the cooling behavior of
neutron stars containing also all hyperons and possibly also a
mixed phase of strange quark matter and hadronic matter, has not been
extensively studied. Two possible reasons for this can be given.
First, the hadronic and mixed equations of state are rather soft
and are continuously being challenged by new data on heavy neutron
stars, such as PSR J1903+0327 which has $M=1.67\pm0.01\textrm{ M}_\odot$
\cite{Freire:2009} and PSR J1614$-$2230 which even has $M=1.97\pm0.04\textrm{ M}_\odot$ \cite{Demorest:2010}. Second, the existence of a mixed phase is being questioned
in view of screening and surface effects \cite{Voskresensky:dual}.
With respect to the first concern, we note that most of the observed
neutron star masses still lie below the maximum mass a
hadronic or mixed equation of state can allow for. According to
Ref.~\cite{Timmers:1996}, most nearby young neutron stars
\cite{Popov:2003} have masses no bigger than $1.4\textrm{ M}_{\odot}$
\cite{Page:2006}. Therefore, at least for the study of the thermal
evolution of these stars, the hadronic or mixed phase can still be
of great importance. In fact, because of the uncertainties in the interaction between particles at extremely high density, it is hard to exactly determine the equation of state at high density. Although the equations of state used in this paper cannot support neutron stars as massive as those reported above, we can still use them for a discussion on medium-mass neutron stars with more complicated compositions. As for the stability of the mixed phase, the arguments are still indecisive. For example, many details of the surface tension, which strongly influences the stability calculation, are still uncertain. Although it is expected that
screening and surface effects diminish the mixed-phase regime, it
is far from certain that its existence can be excluded
\cite{Endo:2011}. The mixed phase can in particular have
significant effects on the cooling behavior, especially for the
heat transport inside the star. However, we show below that its
effect will be less important after the star has become
isothermal, when the thermal evolution is determined by the heat
capacity and neutrino emissivity integrated over the whole volume,
and the inner thermal conductivity no longer plays a role.

The temperature of a neutron star is generally much smaller than
the typical Fermi energy as a consequence of the very high
densities in the star. Therefore, superfluidity may play an
important role. According to BCS theory, fermions can form Cooper pairs at low temperatures via an attractive interaction and thereby
lower the energy of the system. The resulting pairs, which obey
Bose statistics, can form a Bose-Einstein condensate and the
system becomes superfluid. Pair formation changes the
single-particle dispersion around the Fermi surface and
consequently the heat capacity and neutrino emissivity will be
influenced. We find that, without superfluidity, the cooling of
neutron stars is too fast compared with observations. However, by
including the effects of superfluidity we obtain a more realistic
cooling behavior.

The quark phase is usually referred to as an exotic phase of
extremely dense matter, in contrast to the nuclear phase or the
hadronic phase. Other exotic phases have also been proposed, among
which the pion and kaon condensates have attracted much attention
\cite{Ohnishi:2009,Brown:2008}. However, the existence of such
phases inside the neutron star is still an open question. On the
one hand, with such condensates, the equation of state is further
softened and thus the corresponding maximum star mass is reduced
\cite{Potekhin:2011}, which makes such phases less favored when
compared to the observational data of massive stars, as mentioned above. On the other hand, for the cooling process, it was reported that a meson condensate can increase the neutrino emissivity over the typical modified Urca emissivity by several
orders of magnitude \cite{Yakovlev:2001}, but it is still much less
efficient than the direct Urca process. Since in our calculation
the direct Urca process is always present, such enhancement from
the meson condensate has a negligible effect. Besides, the kaon condensation
may even reduce the pressure and cause the star to collapse into a
black hole \cite{Brown:2008}. Therefore, considering all the above
arguments, we do not include such meson condensates in this paper.

The paper is organized as follows, we first introduce in Sec. II our
theoretical framework which includes the relativistic stellar
structure, the unified mean-field model describing the state of matter
inside the star, and the thermal evolution equations. In Sec. III, we then present the numerical results for the cooling curves
of neutron stars with the nuclear, hadronic, and mixed equations
of state. Due to the efficiency of the direct Urca process the
cooling is seen to be too fast compared to observations. Therefore
superfluidity is included in Sec. IV and, as a consequence, the
cooling is slowed down and we find a much better agreement with
observations.

\section{Theoretical framework}
At the high densities of importance to neutron stars the neutrons,
protons, and electrons can be considered to be highly degenerate.
In particular, the typical Fermi temperature of the nucleons is
about $10^{12}\textrm{ K}$  while the temperature of a
proto-neutron star is only $10^{10}\textrm{ K}$. This provides a
great theoretical advantage since it makes it possible to decouple the
calculation of the stellar structure, the nuclear model, and the
thermal evolution of the star. In other words, the thermal
evolution of the star depends on the stellar structure, which in
turn is constructed from the nuclear model. The nuclear model will
be solved in the zero-temperature limit to give the equation of
state. The stellar structure is then obtained from Einstein's
equations using this equation of state.

Although many neutron stars have fast rotation rates and strong
magnetic fields, due to the complexity and richness of these
phenomenon they are beyond our present discussion. In fact, the effect of rotation should be quite small for most of the neutron stars except for millisecond pulsars. Therefore, Einstein's equations are solved for static stars without magnetic
fields. Furthermore, the star interior is approximated by a
perfect fluid, i.e., it is hydrostatic, which is valid as long as
the energy transported by heat conduction is negligible compared
to the total energy. Subsequently, to discuss the thermal
evolution, some properties of the stellar matter, such as the heat
capacities, conductivities, and emissivities, can be obtained by
considering a perturbation of the Fermi surfaces of the various
particles. In the following, we deal with these aspects of our
theoretical framework separately.

\subsection{The general relativistic profile of a compact star}

The space-time metric for a non-rotating, spherically symmetric
star can be written as
\begin{equation}\label{metric}
ds^2=e^{2\Phi(r)}c^2dt^2-e^{2\lambda(r)}dr^2-r^2(d\theta^2+\sin^2\theta d\phi^2),
\end{equation}
where $\Phi(r)$ is the metric function related to the
gravitational redshift, $\exp{[-2\lambda(r)]}=1-2Gm(r)/c^2r$
with $m(r)=\int_0^r4\pi r'^2\rho(r')dr'/c^2$ the gravitational mass
enclosed within the sphere of radius $r$, $G$ is the gravitational
constant, and $c$ is the speed of light. As a consequence of the
above metric, Einstein's equations for the case of a non-rotating,
hydrostatic, and spherically symmetric star reduce to the
Tolman-Oppenheimer-Volkoff (TOV) equations, namely
\begin{align}
\frac{d\Phi(r)}{dr}&=-\frac{1}{\rho(r)+p(r)}\frac{dp(r)}{dr},\\
\frac{dp(r)}{dr}&=-\frac{[\rho(r)+p(r)]G[m(r)+4\pi
r^3p(r)/c^2]}{c^2r^2\left[1-\frac{2Gm(r)}{c^2r}\right]},
\end{align}
where $p(r)$ and $\rho(r)$ are the pressure and energy density at
radius $r$, respectively. Given an equation of state $p(\rho)$, the TOV
equations can be numerically integrated to provide the structure
of the star. The integration starts at $r=0$ with a given pressure
$p(0)=p_c$ until $p(R)=0$ is reached, which defines the radius $R$
of the star. Outside the star, $r>R$, the metric reduces to the
Schwarzschild form,
$\exp{[2\Phi(r)]}=\exp{[-2\lambda(r)]}=1-2GM/c^2r$, where
$M=m(R)$ is the total gravitational mass of the star. Note that
with the help of a local Lorenz transformation we are always able
to apply a locally inertial coordinate system \cite{Weinberg:1972}
such that the calculation of the equation of state is performed in
a homogeneously flat background. Nevertheless, the TOV equations
take into account the effects of general relativity on the
structure of the star. However, to calculate this structure we
need to supply a realistic equation of state for the matter inside
a neutron star.

\subsection{Effective model for matter in neutron stars}

To obtain an equation of state for stellar matter, we apply an
effective nuclear model which includes all relevant kinds of
baryons. The interactions between the baryons are mediated by
three types of effective meson fields, namely the scalar meson
$\sigma$, the vector meson $\omega_\mu$ and the isovector meson
$\vec{\rho}_\mu$. The scalar field is coupled to the derivatives
of the baryon field as shown in the following Lagrangian density
\cite{Glendenning:1992}:
\begin{align}
\mathcal{L}=&\sum_{b}\left[\left(1+\frac{g_{\sigma b}\sigma}{m_bc^2}\right)\bar{\psi}_b\bigg(i\hbar c\gamma_\mu\partial^\mu-g_{\omega b}\gamma_\mu\omega^\mu\right.\nonumber\\
&\quad\left.\left.-\frac{1}{2}g_{\rho b}\gamma_\mu\vec{\tau}\cdot\vec{\rho}^\mu\right)\psi_b-m_bc^2\bar{\psi}_b\psi_b\right]+\frac{\hbar\partial_\mu\sigma\partial^\mu\sigma}{2c^3}\nonumber\\
&-\frac{m^2_\sigma\sigma^2}{2\hbar c}-\frac{\hbar\omega_{\mu\nu}\omega^{\mu\nu}}{4c^3}+\frac{m_\omega^2\omega_\mu\omega^\mu}{2\hbar c}-\frac{\hbar\vec{\rho}_{\mu\nu}\vec{\rho}^{\mu\nu}}{4c^3}\nonumber\\
&+\frac{m_\rho^2\vec{\rho}_\mu\vec{\rho}^\mu}{2\hbar c}+\sum_l\bar{\psi}_l(i\hbar c\gamma_\mu\partial^\mu-m_lc^2)\psi_l,
\end{align}
where $c\partial_0=\partial/\partial t$, the summation over $b$
and $l$ is over all contributing baryon fields $\psi_b$ and lepton
fields $\psi_l$, and in the meson kinetic terms
$\omega_{\mu\nu}\equiv\partial_\mu\omega_\nu-\partial_\nu\omega_\mu$
and
$\vec{\rho}_{\mu\nu}\equiv\partial_\mu\vec{\rho}_\nu-\partial_\nu\vec{\rho}_\mu$
are the antisymmetric field strengths. The derivative coupling of
the $\sigma$ field was first introduced by Zimanyi and Moszkowski
\cite{Zimanyi:1990} to remove the problem of a too small or even
negative reduced baryon mass (commonly referred to as the
effective mass, however, this name is used in this paper only for
the effective fermion mass on the Fermi surface) at high density
in the standard Walecka model \cite{Serot:1986}.

This model can be solved within the mean-field approximation. In
this approximation all the meson fields are replaced by their
ground-state expectation values, which are constants in space-time
and their spatial components are zero because the system is
assumed to be homogeneous and isotropic. Also the charged
components of the isospin vector $\rho^+$ and $\rho^-$, whose
sources are the off-diagonal currents of the baryon fields, are
zero. Therefore, only the three constant fields $\sigma$,
$\omega^0$ and $\rho_3^0$ survive in the Euler-Lagrange equations.
In the following, we simply denote the last two as $\omega$ and
$\rho$. Furthermore, the constant scalar field $\sigma$ can be
absorbed into the baryon field and the reduced mass by a
rescaling, namely $\Psi_b=\sqrt{1+\frac{g_{\sigma
b}\sigma}{m_bc^2}}\psi_b$ and
$\tilde{m}_b(\sigma)=m_b/(1+\frac{g_{\sigma b}\sigma}{m_bc^2})$.
Notice that $\tilde{m}_b$ is positive definite and only approaches
zero as $\sigma\rightarrow\infty$.

In the mean-field approximation the Lagrangian gives rise to the
following field equations:
\begin{align}
&\left[i\hbar c\gamma^\mu\partial_\mu-g_{\omega b}\gamma_0\omega-\frac{1}{2}g_{\rho b}\gamma_0\tau_3\rho-\tilde{m}_bc^2\right]\Psi_b=0,\label{eqnb}\\
&\sum_b\frac{g_{\sigma b}\langle\bar{\Psi}_b\Psi_b\rangle}{\left(1+\frac{g_{\sigma b}\sigma}{m_bc^2}\right)^2}-\frac{m_\sigma^2\sigma}{\hbar c}=0,\label{eqns}\\
&\sum_bg_{\omega b}\langle\Psi^\dagger_b\Psi_b\rangle-\frac{m_\omega^2\omega}{\hbar c}=0,\label{eqno}\\
&\sum_bg_{\rho b}\frac{1}{2}\langle\Psi^\dagger_b\tau_3\Psi_b\rangle-\frac{m_\rho^2\rho}{\hbar c}=0,\label{eqnr}
\end{align}
where $\tau_3/2$ gives the expectation value of the third
component of the baryon isospin $I_{3b}$. The equation of motion
of the baryons gives the following energy eigenvalue
\begin{equation}\label{baryon}
\mu_b=\epsilon_b=g_{\omega b}\omega+g_{\rho b}I_{3b}\rho+\sqrt{\hbar^2c^2k_b^2+\tilde{m}_b^2c^4}.
\end{equation}
From the above it is clear that the baryon $b$ only exists when
$\mu_b-(g_{\omega b}\omega+g_{\rho b}I_{3b}\rho)>\tilde{m}_bc^2$. The
equation of motion of the leptons are simply the free Dirac
equations and are not listed here. Thus, the leptons obey the
simple relations of a free relativistic Fermi gas:
\begin{equation}\label{lepton}
n_l=f_l\frac{k_l^3}{6\pi^2},\quad \mu_l=\epsilon_l=\sqrt{\hbar^2c^2k_l^2+m_l^2c^4}.
\end{equation}
For the expectation value of the baryon field we have
\begin{align}
\langle\bar{\Psi}_b\Psi_b\rangle&=\frac{f_b}{2\pi^2}\int_0^{k_b}\frac{\tilde{m}_bc^2k^2dk}{\sqrt{\hbar^2c^2k^2+\tilde{m}_b^2c^4}}\nonumber\\
&=\frac{f_b}{2\pi^2}\frac{\tilde{m}_bc}{2\hbar^3}\bigg[\hbar k_b\sqrt{\hbar^2k_b^2+\tilde{m}_b^2c^2}\nonumber\\
&\quad\left.-\tilde{m}_b^2c^2\log\left(\frac{\hbar k_b}{\tilde{m}_bc}+\sqrt{1+\frac{\hbar^2k_b^2}{\tilde{m}_b^2c^2}}\right)\right],
\end{align}
and
\begin{equation}
\langle\Psi^\dagger_b\Psi_b\rangle=n_b=f_b\frac{k_b^3}{6\pi^2}.
\end{equation}
In the above equations $f_i=2J_i+1$ is the particle degeneracy, while
$k_i$ and $\epsilon_i$ are the Fermi momentum and Fermi energy,
respectively.

Since the whole star is considered to be in equilibrium, the
various processes, such as the direct Urca process in
Eq.~(\ref{DUrca}) and those involving the hyperons in
Eq.~(\ref{lambda}), impose relations between the chemical
potentials of the particles. During these processes the baryon
number and electric charge are conserved, which gives rise to the
two chemical potentials $\mu_B$ and $\mu_Q$ for baryon number and
electric charge, respectively. The chemical potential for an
arbitrary particle with baryon number $B_i$ and charge $Q_i$, such
as a baryon, lepton or quark, can be written as
$\mu_i=B_i\mu_B+Q_i\mu_Q$. The particle data for all the fermions
used in this paper are listed in Table.~\ref{data}.
\begin{table}
\caption{Particle Data \cite{PDG:2010}. Here, $M$ is the particle mass,
$B$ is the baryon number, $J$ is the spin, $I_3$ is the
3-component of the isospin, and $Q$ is the charge.}
\begin{tabular}{|c||c|c|c|c|c|c|}\hline
& particle & $M$/MeV & $B$ & $J$ & $I_3$ & $Q/e$ \\\hline\hline
& $p$ & $938.27$ & $1$ & $1/2$ & $1/2$ & $+1$ \\
& $n$ & $939.57$ & $1$ & $1/2$ & $-1/2$ & $0$ \\
& $\Lambda^0$ & $1115.7$ & $1$ & $1/2$ & $0$ & $0$ \\
& $\Sigma^+$ & $1189.4$ & $1$ & $1/2$ & $1$ & $+1$ \\
& $\Sigma^0$ & $1192.6$ & $1$ & $1/2$ & $0$ & $0$ \\
& $\Sigma^-$ & $1197.4$ & $1$ & $1/2$ & $-1$ & $-1$ \\
Baryon & $\Delta^{++}$ & $1232$ & $1$ & $3/2$ & $+3/2$ & $+2$ \\
& $\Delta^+$ & $1232$ & $1$ & $3/2$ & $+1/2$ & $+1$ \\
& $\Delta^0$ & $1232$ & $1$ & $3/2$ & $-1/2$ & $0$ \\
& $\Delta^-$ & $1232$ & $1$ & $3/2$ & $-3/2$ & $-1$ \\
& $\Xi^0$ & $1315$ & $1$ & $1/2$ & $+1/2$ & $0$ \\
& $\Xi^-$ & $1322$ & $1$ & $1/2$ & $-1/2$ & $-1$ \\
& $\Omega^-$ & $1672$ & $1$ & $3/2$ & $0$ & $-1$ \\\hline
& $e^-$ & $0.511$ & $0$ & $1/2$ & $0$ & $-1$ \\
\raisebox{1.5ex}[0cm][0cm]{Lepton}
& $\mu^-$ & $105.7$ & $0$ & $1/2$ & $0$ & $-1$ \\\hline
& $u$ & $2.4$ & $1/3$ & $1/2$ & $+1/2$ & $+2/3$ \\
Quark & $d$ & $4.9$ & $1/3$ & $1/2$ & $-1/2$ & $-1/3$ \\
& $s$ & $105$ & $1/3$ & $1/2$ & $0$ & $-1/3$ \\\hline
\end{tabular}\label{data}
\end{table}

The coupling constants $g_{\sigma b}$, $g_{\omega b}$ and $g_{\rho
b}$ for nucleons can be fitted with data of the interactions at
densities near the nuclear saturation point. However, for hyperons
the values are not known. Here the coupling constants are simply
chosen to be equal among the different baryons
\cite{Glendenning:1992}. Their values are
$(g_\sigma/m_\sigma)^2=7.487\textrm{ fm}^2$,
$(g_\omega/m_\omega)^2=2.615\textrm{ fm}^2$ and
$(g_\rho/m_\rho)^2=4.774\textrm{ fm}^2$. Notice that the ratio of
the coupling constant and the corresponding meson mass is enough
to solve this model, because we can always rescale the meson
fields with their masses, so the meson mass will not appear in
Eqs.~(\ref{eqnb}-\ref{eqnr}).

\begin{figure}
\resizebox{1\linewidth}{!}{\includegraphics*{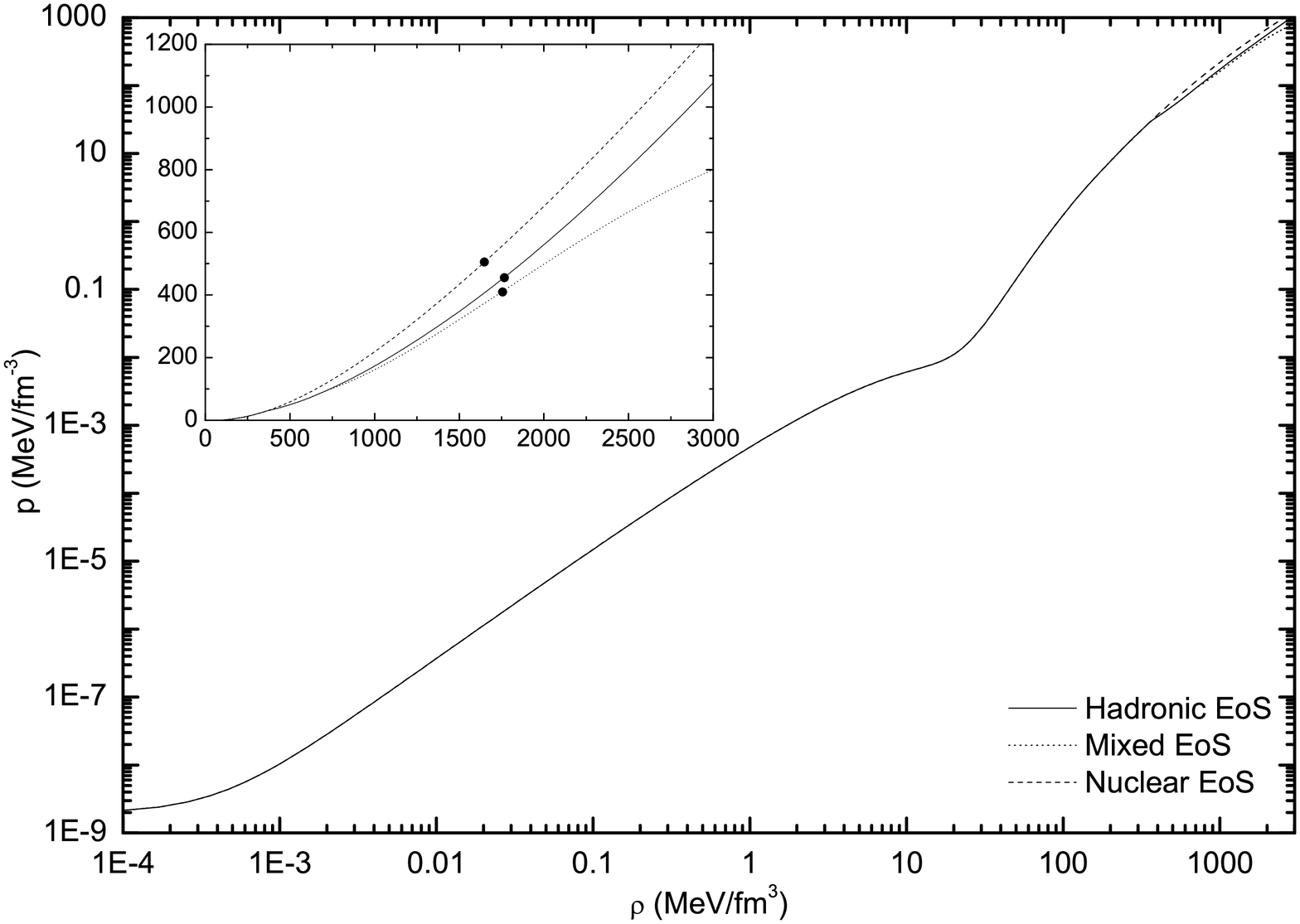}} \caption{The equation of state of the various matter phases. The difference only appears at high density as is shown more clearly in the inset, where the dots on the curves indicate the central densities and pressures of the corresponding maximum mass stars.}\label{EoS}
\end{figure}

\begin{figure}
\resizebox{1\linewidth}{!}{\includegraphics*{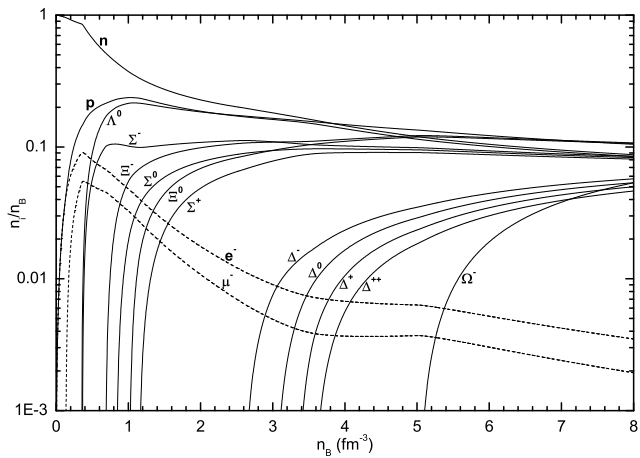}} \caption{The particle densities $n_i/n_B$ versus baryon density $n_B$ for hadronic matter.}\label{hadronic}
\end{figure}

\begin{figure}
\resizebox{1\linewidth}{!}{\includegraphics*{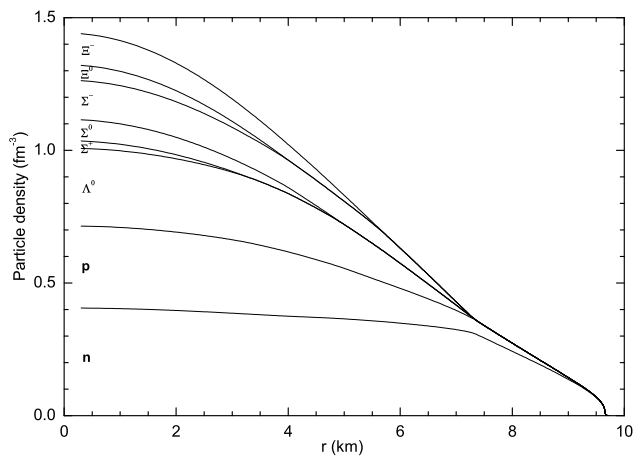}} \caption{The particle composition of the maximum mass neutron star with the hadronic equation of state.}\label{hadronicrn}
\end{figure}

Given the chemical potentials $\mu_B$ and $\mu_Q$ as the input,
together with the neutrality condition $\sum_in_iQ_i=0$,
Eqs.~(\ref{eqnb}-\ref{eqnr}) form a set of self-consistent
nonlinear equations which can be solved numerically. With the
Fermi momenta $k_i$ of the particles as the output, we can
calculate the energy density and the pressure of the hadronic
matter as:
\begin{align}
\rho_H&=\sum_b\frac{f_bc}{16\pi^2\hbar^3}\bigg[\hbar k_b\sqrt{\hbar^2k_b^2+\tilde{m}_b^2c^2}(2\hbar^2k_b^2+\tilde{m}_b^2c^2)\nonumber\\
&\qquad\left.-\tilde{m}_b^4c^4\log\left(\frac{\hbar k_b}{\tilde{m}_bc}+\sqrt{1+\frac{\hbar^2k_b^2}{\tilde{m}_b^2c^2}}\right)\right]\nonumber\\
&\qquad+\frac{1}{2\hbar c}(m^2_\sigma\sigma^2+m^2_\omega\omega^2+m^2_\rho\rho^2)\nonumber\\
&\quad+\sum_l\frac{f_lc}{16\pi^2\hbar^3}\bigg[\hbar k_l\sqrt{\hbar^2k_l^2+m_l^2c^2}(2\hbar^2k_l^2+m_l^2c^2)\nonumber\\
&\qquad\left.-m_l^4c^4\log\left(\frac{\hbar k_l}{m_lc}+\sqrt{1+\frac{\hbar^2k_l^2}{m_l^2c^2}}\right)\right],\label{hadronrho}
\end{align}
\begin{align}
p_H&=\sum_b\frac{f_bc}{48\pi^2\hbar^3}\bigg[\hbar k_b\sqrt{\hbar^2k_b^2+\tilde{m}_b^2c^2}(2\hbar^2k_b^2-3\tilde{m}_b^2c^2)\nonumber\\
&\qquad\left.+3\tilde{m}_b^4c^4\log\left(\frac{\hbar k_b}{\tilde{m}_bc}+\sqrt{1+\frac{\hbar^2k_b^2}{\tilde{m}_b^2c^2}}\right)\right]\nonumber\\
&\quad-\frac{1}{2\hbar c}(m^2_\sigma\sigma^2-m^2_\omega\omega^2-m^2_\rho\rho^2)\nonumber\\
&+\sum_l\frac{f_lc}{48\pi^2\hbar^3}\bigg[\hbar k_l\sqrt{\hbar^2k_l^2+m_l^2c^2}(2\hbar^2k_l^2-3m_l^2c^2)\nonumber\\
&\qquad\left.+3m_l^4c^4\log\left(\frac{\hbar k_l}{m_lc}+\sqrt{1+\frac{\hbar^2k_l^2}{m_l^2c^2}}\right)\right].\label{hadronp}
\end{align}
The relation between the pressure $p$ and the energy density
$\rho$ gives the equation of state, which is shown in
Fig.~\ref{EoS}, and is subsequently used in the
TOV equations to get the stellar structure. The particle densities as a function of baryon density are shown in Fig.~\ref{hadronic}. Fig.~\ref{hadronicrn}
gives the particle composition inside the maximum mass star with
the hadronic equation of state ($M=1.523\textrm{ M}_\odot$,
$R=9.72\textrm{ km}$). Remember that the model is used to
calculate the equation of state for the whole range of the
density, however, this description may not be appropriate near the
star surface, i.e., the crust. Nevertheless, the outer crust only
affects low mass stars significantly, while we are mainly
concerned with maximum mass stars. Furthermore, in view of the
cooling behavior, the outer crust has negligible neutrino
emissivity compared to the direct Urca process in the core.

Compared to the hadronic phase, the quark phase is more easily
described. Quark matter is treated as a free Fermi gas, whereby we
assume that asymptotic freedom has taken effect at the very high
densities in the center of the star. Thus Eq.~(\ref{lepton}) is
also valid for quarks, where the lepton chemical potential $\mu_l$
is substituted with the quark chemical potential $\mu_q$ and the
degeneracy becomes $f_q=3\times(2J_q+1)=6$ because of the extra
color degrees of freedom. The fact that the quark phase has a
different vacuum than the hadronic phase, which has a nonzero expectation value of the gluon field, can be taken into account
by the so-called bag model. This model adds a constant shift
called the bag constant $B_c$ to the pressure and energy density of
the quarks, such that
\begin{align}
\rho_q&=\sum_q\frac{f_q}{16c\pi^2\hbar^3}\bigg[\hbar k_q\sqrt{\hbar^2k_q^2+m_q^2c^2}(2\hbar^2k_q^2+m_q^2c^2)\nonumber\\
&\quad\left.-m_q^4c^4\log\left(\frac{\hbar k_q}{m_qc}+\sqrt{1+\frac{\hbar^2k_q^2}{m_q^2c^2}}\right)\right]\nonumber\\
&+\sum_l\frac{f_l}{16c\pi^2\hbar^3}\bigg[\hbar k_l\sqrt{\hbar^2k_l^2+m_l^2c^2}(2\hbar^2k_l^2+m_l^2c^2)\nonumber\\
&\quad\left.-m_l^4c^4\log\left(\frac{\hbar k_l}{m_lc}+\sqrt{1+\frac{\hbar^2k_l^2}{m_l^2c^2}}\right)\right]+B_c,\label{quarkrho}\\
p_q&=\sum_q\frac{f_q}{48c\pi^2\hbar^3}\bigg[\hbar k_q\sqrt{\hbar^2k_q^2+m_q^2c^2}(2\hbar^2k_q^2-3m_q^2c^2)\nonumber\\
&\quad\left.+3m_q^4c^4\log\left(\frac{\hbar k_q}{m_qc}+\sqrt{1+\frac{\hbar^2k_q^2}{m_q^2c^2}}\right)\right]\nonumber\\
&+\sum_l\frac{f_l}{48c\pi^2\hbar^3}\bigg[\hbar k_l\sqrt{\hbar^2k_l^2+m_l^2c^2}(2\hbar^2k_l^2-3m_l^2c^2)\nonumber\\
&\quad\left.+3m_l^4c^4\log\left(\frac{\hbar k_l}{m_lc}+\sqrt{1+\frac{\hbar^2k_l^2}{m_l^2c^2}}\right)\right]-B_c,\label{quarkp}
\end{align}
where the leptons still contribute because the unequal masses of
the $u$, $d$ and $s$ quarks result in unequal densities of these
flavors, such that even in the pure quark phase charge neutrality
cannot be satisfied without leptons.

\begin{figure}
\resizebox{3.5in}{!}{\includegraphics*{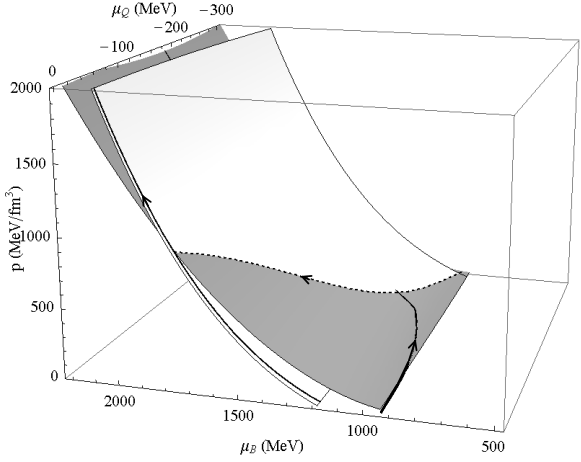}} \caption{The pressure surface of the hadronic phase (dark gray) and the quark phase (light gray). The thick solid lines are the neutrality curves in each phase. The thick dashed curve is the intersection of the two pressure surfaces and shows when the two phases are in equilibrium. The arrows indicate how $p$ and $\mu_Q$ change with increasing $\mu_B$.}\label{pressure}
\end{figure}

\begin{figure}
\resizebox{1\linewidth}{!}{\includegraphics*{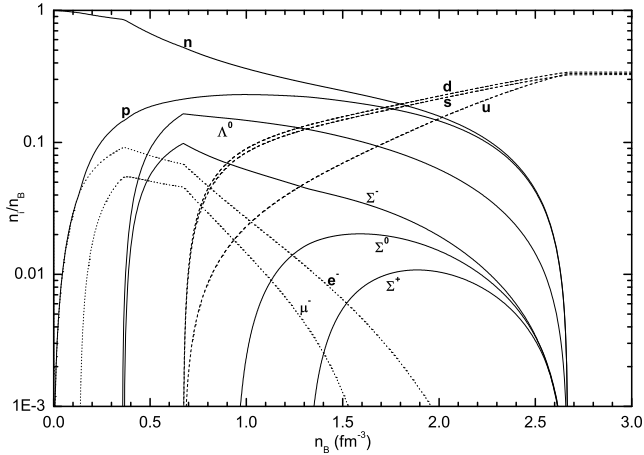}} \caption{The particle densities $n_i/n_B$ versus baryon density $n_B$ for the mixed phase case. The densities of the quarks are expressed in terms of their baryon densities instead of their number densities.}\label{hadronicmix}
\end{figure}

\begin{figure}
\resizebox{1\linewidth}{!}{\includegraphics*{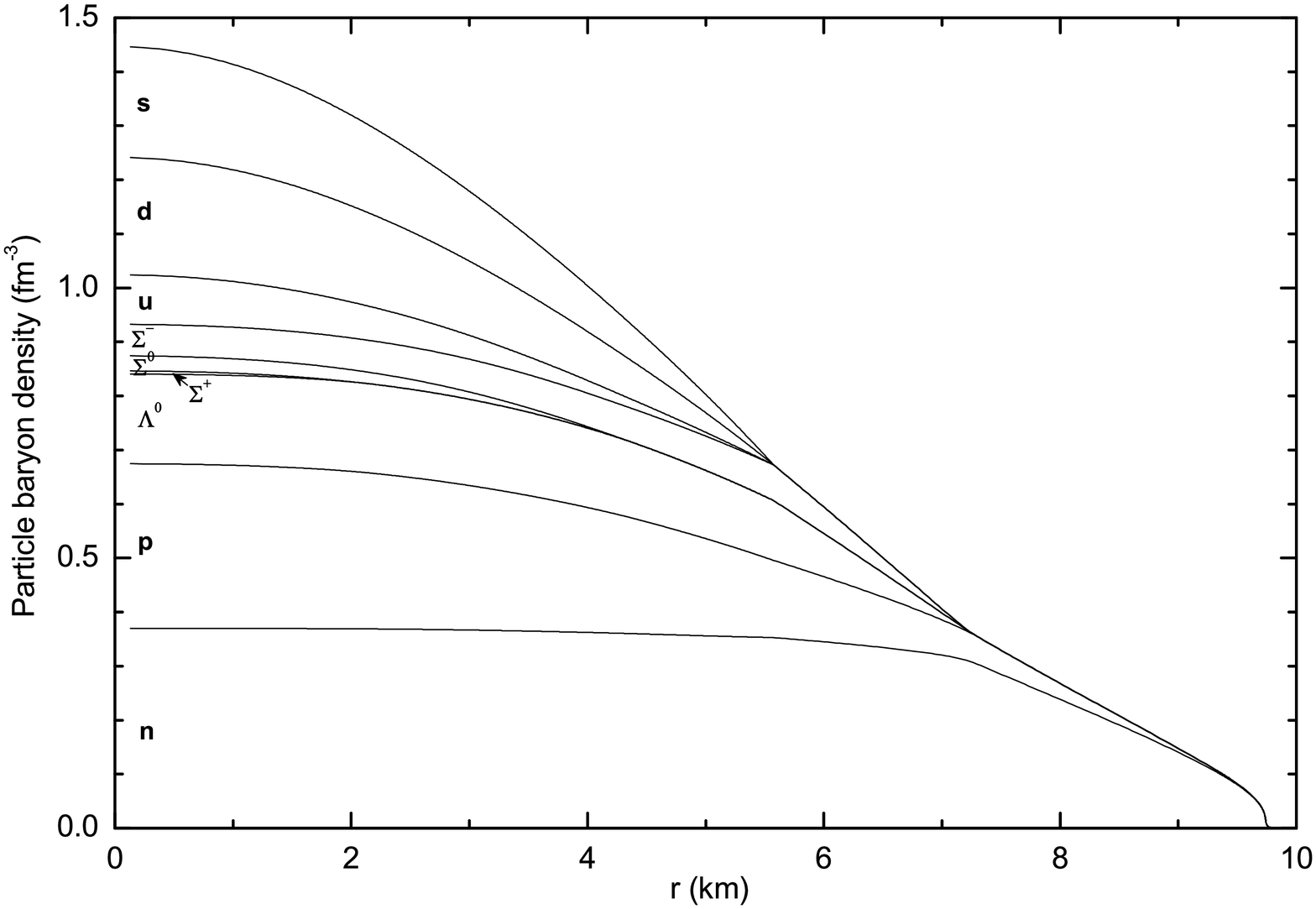}} \caption{The particle baryon densities inside the maximum mass neutron star with the mixed equation of state. Note that the quarks have baryon number $1/3$.}\label{hadronicmixrn}
\end{figure}

As discussed in the introduction, there could also be a mixed
phase between the hadronic phase and the quark phase. These two
phases are taken to be in equilibrium, i.e., the two phases have
the same temperature (both set to zero here), pressure, and
chemical potentials. Charge neutrality then determines the volume
fraction of these two phases. For the three possible phases
(hadronic, quark, and mixed) the system will be in the one with
the highest pressure, or equivalently, the lowest grand potential,
which determines the phase transition behavior. According to Eq.
(\ref{quarkp}), the bag constant determines the scale at which
deconfinement sets in. In other words, the larger the bag
constant, the later the quark phase sets in with increasing
density. Throughout this paper the bag constant is taken to be
$B_c=230\textrm{ MeV}/\textrm{fm}^3$ unless indicated specifically.
In Fig. \ref{pressure} it is shown how the phase transition takes
place between the two phases. Similar to Fig. \ref{hadronic} for
the hadronic phase, the various particle densities for the mixed
phase are shown in Fig. \ref{hadronicmix} as a function of baryon
density. Fig. \ref{hadronicmixrn} shows the particle composition
inside the maximum mass star with the mixed equation of
state ($M=1.479\textrm{ M}_\odot$, $R=9.81\textrm{ km}$).

\subsection{Thermal evolution equations}

\begin{figure}
\resizebox{1\linewidth}{!}{\includegraphics*{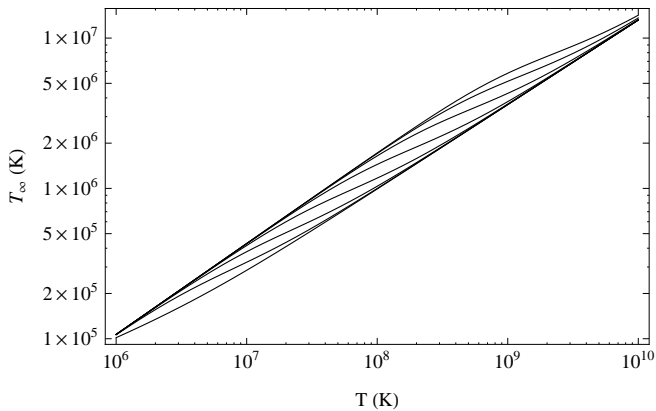}} \caption{The $T_\infty$-$T$ relation for different numbers of light elements parameterized by $\eta=g_{14}^2\Delta M_l/M$, where $\Delta M_l$ is the mass of the light elements in the envelop and $g_{14}$ is the surface gravity in unit of $10^{14}\textrm{ cm}/\textrm{s}^2$ \cite{Potekhin:1997}, the different curves correspond to $\log\eta=-\infty$, $-16$, $-14$, $-12$, $-10$, $-8$, $-6$ and $0$ from bottom to top, respectively. These results are for the maximum mass neutron star with the hadronic equation of state, where $g_{14}=2.918$.}\label{TT}
\end{figure}

The equations governing the thermal evolution of a spherically
symmetric star, given by the metric in Eq. (\ref{metric}), are
\cite{Thorne:1977}:
\begin{align}
c_V\frac{\partial(Te^\Phi)}{\partial t}&=-\frac{e^{-\lambda}}{4\pi r^2}\frac{\partial(Le^{2\Phi})}{\partial r}-q_\nu e^{2\Phi}-q_\gamma e^{2\Phi},\label{therm1}\\
\kappa\frac{\partial(Te^{\Phi})}{\partial r}&=-\frac{Le^{\lambda+\Phi}}{4\pi r^2},\label{therm2}
\end{align}
where $c_V$ is the specific heat capacity at constant volume,
$\kappa$ is the thermal conductivity, $q_\nu$ and $q_\gamma$ are
the neutrino and photon emissivity, respectively. The photon
emissivity $q_\gamma$ is only nonzero on the surface $r=R$. It is
convenient to define the redshifted temperature
$\tilde{T}=Te^{\Phi}$ inside the star. Similarly, $Le^{2\Phi}$
represents the redshifted luminosity corresponding to the heat
current. In order to compare with observations, the photon
luminosity is of great importance. Here it is assumed to obey the
black body radiation law $L_\gamma=4\pi R^2q_\gamma=4\pi\sigma
R^2T_s^4$, where $\sigma=5.67\times10^{-5}\textrm{ erg}/\textrm{cm}^2\textrm{K}^4\textrm{s}$ is the
Stefan-Boltzmann constant and $T_s$ is the surface temperature.
According to general relativity, an observer at infinity will
measure the gravitationally red-shifted temperature and
luminosity. The observational quantities are thus $T_\infty=T_se^{\Phi(R)}$ and
$L_\infty=L_\gamma e^{2\Phi(R)}$. As
is generally accepted, a very thin layer of the outer crust of the
neutron star will act as a thermal insulator causing the
temperature at the surface to be much lower than inside the star.
The relation between the surface temperature and the inner
temperature depends on the chemical composition of this envelop.
We will simply locate such a layer at the surface, neglecting its
thickness, and adopt the $T_\infty$-$T$ relation given by Potekhin {\it et al.}
\cite{Potekhin:1997}, as shown in Fig.~\ref{TT}, and set $T$ to be
the inside temperature at $r=R$. Note that the surface temperature $T_s$ introduced above is the outside temperature at $r=R$.

The parameters $c_V$, $\kappa$, $q_\nu$ and $q_\gamma$ need to be
determined to solve the thermal evolution equations, Eqs.
(\ref{therm1}) and (\ref{therm2}). The specific heat $c_V$ can simply be
obtained by summing up all the contributions of the different
fermions in the model. In the low-temperature approximation,
Fermi-liquid theory gives
\begin{equation}
c_{Vi}=\frac{m_i^*k_i}{3\hbar^2}k_B^2T,
\end{equation}
where $k_B$ is Boltzmann's constant, $m_i^*=\hbar k_i/v_i$ is the
effective mass on the Fermi surface, and $v_i=d\epsilon_i/(\hbar
dk)|_{k=k_i}$ is the Fermi velocity of quasi-particle $i$. For
leptons the effective mass is easily obtained since they are
described as a free Fermi gas, i.e.,
$m^*_l=\sqrt{\hbar^2k_l^2/c^2+m_l^2}=\mu_l/c^2$. Similarly for quarks, $m_q^*=\mu_q/c^2$. For baryons the
effective mass follows from the dispersion relation in
Eq.~(\ref{baryon})
\begin{equation}
v_b=\frac{d\epsilon_b(k)}{\hbar dk}\bigg|_{k=k_b}=\frac{\hbar k_b}{\sqrt{\hbar^2k_b^2/c^2+\tilde{m}_b^2}},
\end{equation}
such that $m^*_b=\sqrt{\hbar^2k_b^2/c^2+\tilde{m}_b^2}$. The
effect of interactions is, in the mean-field approximation, only
present in the reduced mass $\tilde{m}_b$ of the baryons.

The general expression for the neutrino emissivity has the form
$q_{\nu i}=C_iT^s$, where $C_i$ and $s$ are different constants
for each kind of process. For the direct Urca process we have
$s=6$ and for the modified Urca process $s=8$. The difference in
the exponents shows the inefficiency of the modified Urca process
since the temperature $T$ of a neutron star is much smaller than
the typical Fermi temperature $T_F$. The neutrino emissivities for
the various processes are summarized in Ref.~\cite{Yakovlev:2001}.
For the general baryon direct Urca process, cf.
Eq.~(\ref{DUrcabaryon}), the emissivity is
\begin{equation}
q_{\nu12l}\approx1.207\times10^{25}\frac{\mu_lm^*_{b1}m^*_{b2}}{(1{\textrm{  MeV}})m^2_n}R_{12}T_9^6\Theta_{12l}\textrm{ erg}/\textrm{cm}^{3}\textrm{s},
\end{equation}
where $T_9$ is the temperature in units of $10^9\textrm{ K}$,
$\Theta_{12l}$ is $1$ only if the Fermi momenta of the two baryons
$b_1$, $b_2$ and the lepton $l$ can form a triangle, otherwise
$\Theta_{12l}=0$. The coefficients $R_{12}$ vary depending on the
baryons involved in the process. In our mean-field model the
maximum mass star contains massive hyperons up to $\Xi$, such that
all possible direct Urca processes given in
Ref.~\cite{Prakash:1992} need to be included, as listed in Table~\ref{RDUrca}.
\begin{table}
\caption{The coefficients $R_{12}$ for different direct Urca processes \cite{Prakash:1992}, denoted by the baryons involved in each process.}
\begin{tabular}{|c|c|c|c|c|c|c|c|c|}\hline
$np$&$\Lambda^0p$&$\Sigma^-n$&$\Sigma^-\Lambda^0$&$\Sigma^-\Sigma^0$&$\Xi^-\Lambda^0$&$\Xi^-\Sigma^0$&$\Xi^0\Sigma^+$&$\Xi^-\Xi^0$\\\hline
$1$&$0.0394$&$0.0125$&$0.2055$&$0.6052$&$0.0175$&$0.0282$&$0.0564$&$0.2218$\\\hline
\end{tabular}\label{RDUrca}
\end{table}
For the quark direct Urca emissivities, we refer to
Ref.~\cite{Iwamoto:1982}:
\begin{align}
q_{\nu udl}&\approx4.773\times10^{18}\frac{\hbar^2c^2[(k_u+k_l)^2-k_d^2]\mu_d}{1\textrm{  MeV}^3}T_9^6\textrm{ erg}/\textrm{cm}^{3}\textrm{s},\nonumber\\
q_{\nu usl}&\approx2.552\times10^{17}\frac{\hbar^2c^2[(k_u+k_l)^2-k_s^2]\mu_s}{1{\textrm{  MeV}^3}}T_9^6\textrm{ erg}/\textrm{cm}^{3}\textrm{s},
\end{align}
where the efficiency of the strange quark channel is reduced
because it changes strangeness. Since the direct Urca channels are
open, it is not necessary to consider other neutrino-emission
processes.

In the present model, the thermal conductivity $\kappa$ cannot be
easily obtained since it strongly depends on the interactions
between the particles in the complicated hadronic phase. For the mixed case, the
interface between the phases may introduce even more uncertainty.
Therefore, the thermal relaxation period inside the star is not
considered here and the calculation simply starts when the star
has become isothermal, as indicated in the introduction. Because
of this simplification the interesting early stage behavior of the
cooling cannot be discussed. This disadvantage may also be one of
the reasons why such hadronic models are not extensively studied.
However, the typical thermal relaxation time is usually less than
a century and most of the available observational data is of the
later stages, i.e., after the star has become isothermal, such
that the results obtained after the relaxation period has ended
are still useful. Consequently, the star will be treated as
an isothermal object with a constant temperature $\tilde{T}$ and the
equations are rather simplified. Instead of two coupled partial
differential equations in Eqs.~(\ref{therm1}) and (\ref{therm2}), we
have now only one ordinary differential equation:
\begin{equation}\label{therm}
C_V\frac{d\tilde{T}}{dt}=-Q_\nu-L_\infty,
\end{equation}
where the capital letters represent the integrated parameters over
the volume of the star, namely
\begin{align}\label{CV}
C_V&=4\pi\tilde{T}\sum_i\int_0^R\frac{\bar{c}_{Vi}(r)e^{\lambda(r)}r^2}{e^{\Phi(r)}}dr,\\\label{Qnv}
Q_\nu&=4\pi\tilde{T}^s\sum_i\int_0^R\frac{\bar{q}_{\nu i}(r)e^{\lambda(r)}r^2}{e^{(s-2)\Phi(r)}}dr,
\end{align}
where the $T$-dependent parts are taken outside of the radial
integral and the temperature-independent prefactors of $c_{Vi}$ and
$q_{\nu i}$ are denoted with a bar. The factors of $e^{\Phi(r)}$
in the denominators are a consequence of the construction of the
isothermal temperature $\tilde{T}$. Note that the isothermal
temperature does not take into account the temperature decrease
near the surface of the star, which occurs in a very thin layer
close to the surface.

\begin{figure}
\resizebox{1\linewidth}{!}{\includegraphics*{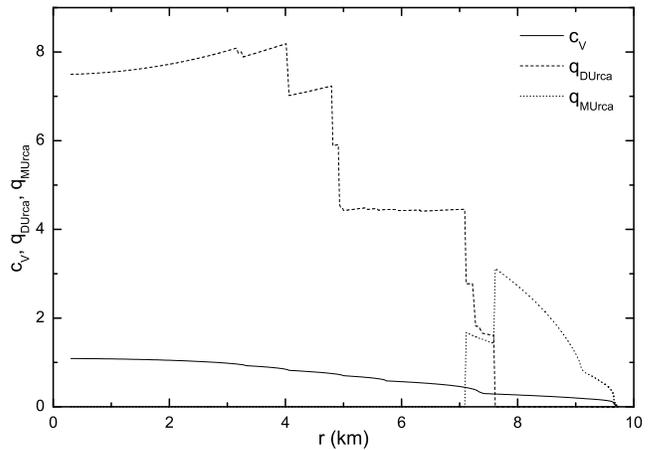}} \caption{The total heat capacity ($c_V$) and the neutrino emissivities of the direct Urca ($q_{\rm DUrca}$) and modified Urca ($q_{\rm MUrca}$) processes inside the neutron star with the hadronic equation of state at $T=10^9\textrm{ K}$. Here, $c_V$ is in units of $10^{21}\textrm{ erg}/\textrm{cm}^3 \textrm{K}$, $q_{\rm DUrca}$ is in units of $10^{27}\textrm{ erg}/\textrm{cm}^3 \textrm{K}$, and $q_{\rm MUrca}$ is in units of $10^{19}\textrm{ erg}/\textrm{cm}^3\textrm{K}$.}\label{hadronictherm}
\end{figure}

\begin{figure}
\resizebox{1\linewidth}{!}{\includegraphics*{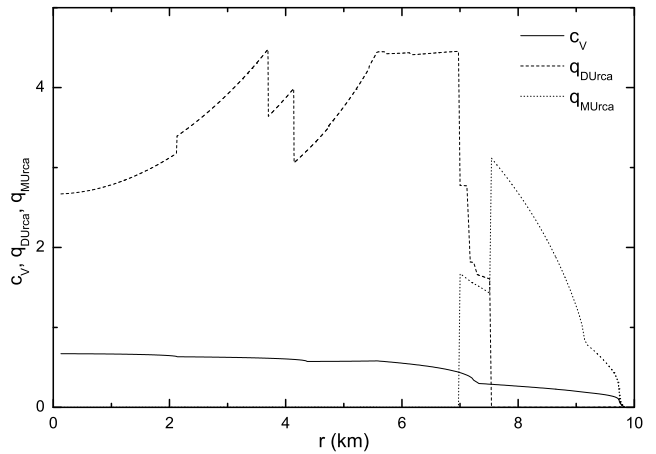}} \caption{The same as in Fig.~\ref{hadronictherm} but now for the mixed equation of state.}\label{hadronicmixtherm}
\end{figure}

The total heat capacity and the neutrino emissivities for both the
direct and modified Urca processes as a function of the radial
coordinate are shown in Figs.~\ref{hadronictherm} and
\ref{hadronicmixtherm} for the case of a maximum mass neutron
star with the hadronic and mixed equations of state,
respectively. Here ``total" means summed over all particles which
contribute, but not integrated over the star volume. The
quantities were obtained at $T=10^9\textrm{ K}$,
but the values at different temperatures can be obtained directly
from the temperature dependence of $c_V$ and $q_\nu$. The
discontinuities in the emissivity are due to the step functions
in the expressions for the emissivities. These step functions are
a consequence of the low-temperature approximation, which does not
take into account any further momentum fluctuations around the
Fermi surfaces. Also $q_{\rm MUrca}$ is calculated only for the nucleon processes shown in
Eq.~(\ref{MUrca}) with the following emissivity
\cite{Yakovlev:2001}:
\begin{align}
q_{\nu Mn}&\approx1.882\times10^{19}\frac{m_n^{*3}m_p^*\hbar^2c^2k_pk_l}{(1\textrm{ MeV})\mu_lm_n^3m_p}T_9^8\textrm{ erg}/\textrm{cm}^{3}\textrm{s},\nonumber\\
q_{\nu Mp}&\approx\left(\frac{m_p^*}{m_n^*}\right)^2\frac{(3k_p+k_l-k_n)^2}{8k_pk_l}\Theta_{Mp}q_{\nu Mn},
\end{align}
where the subscript $Mn$ or $Mp$ means the process is modified by
a bystander neutron $n$ or proton $p$, and $\Theta_{Mp}$ is $1$
only if $3k_p+k_l>k_n$ as required by momentum conservation in the
$Mp$ process. The modified Urca processes are only calculated when
the nucleon direct Urca process is forbidden, since the formulae
are only valid in this case. The small region of overlap where
both $q_{\rm DUrca}$ and $q_{\rm MUrca}$ are nonzero is due to the
different thresholds for the direct Urca processes with electrons
$e^-$ and muons $\mu^-$.

\section{Numerical result}

In this section the three different equations of state for the
hadronic matter, the mixed phase of hadronic and quark matter, and
nuclear matter are compared. Since the maximum masses are not very
large for the last two cases, we concentrate on the star with the
maximum allowable mass for each equation of state. The initial
temperature is always set as $\tilde{T}=10^9 \textrm{ K}$ at
$t=0$, which corresponds to the time at which the isothermal
condition is reached. As mentioned previously, the thermal relaxation period is not
included in our calculation, which is about several decades
according to calculations with nuclear matter
\cite{Yakovlev:2001}. The results are plotted in a logarithmic time scale, such that our results can be shown directly with most of the observational data even though the data include the thermal relaxation era of the neutron stars, that is negligible compared to the age of the stars. However, this makes it hard to draw a comparison with the recent observation of the cooling of the Cassiopeia A neutron star \cite{Heinke:2010}, which provides direct evidence for the cooling process. This neutron star is too young and is believed to have become isothermal quite recently, such that the thermal relaxation period is not negligible.

In order to avoid confusion, our results always show cooling curves starting at $t=1\textrm{ yr}$, when the result has become not very sensitive to the initial conditions. In fact, the effect of the initial temperature is limited to the very beginning. The higher the
initial temperature is, the faster the dependence dies away, which
can be seen clearly from the asymptotic temperature dependence of
the neutrino-emission dominated era
\begin{equation}
\tilde{T}=\left[\frac{(s-2)Q_0}{C_0}t+\tilde{T}(0)^{2-s}\right]^{\frac{1}{2-s}},
\end{equation}
where $C_0$ and $Q_0$ are the constant coefficients of the
$\tilde{T}$-dependent factors in Eqs.~(\ref{CV}) and (\ref{Qnv}). This
asymptotic solution can be easily obtained by dropping the photon
radiation term in Eq.~(\ref{therm}) and keeping only the leading
neutrino-emission process with the smallest exponent $s$. Note
that the exponent $s$ is never smaller than $6$ for any of the
neutrino-emission processes. It is quite clear that the late-time
behavior of the neutrino-emission dominated era is thus completely
determined by the coefficients of the heat capacity and the
neutrino emissivity and not by the initial condition $\tilde{T}(0)$.

\begin{figure}
\resizebox{1\linewidth}{!}{\includegraphics{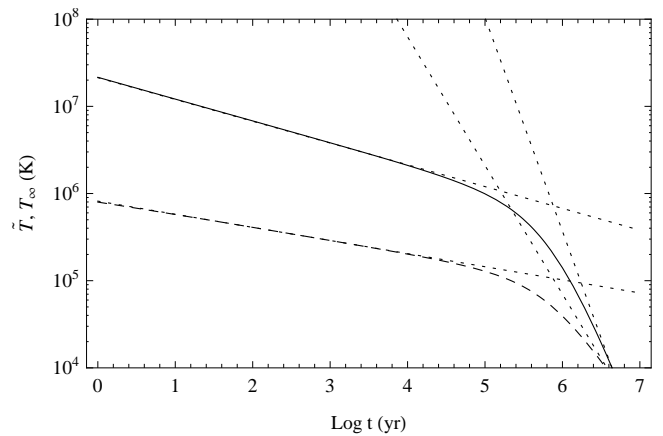}}
\caption{The temperature evolution of the neutron star with the hadronic equation of state. The solid curve represents $\tilde{T}$ while the dashed curve represents $T_\infty$, with the dotted straight lines indicating the asymptotic behavior.}\label{hadronic-T}
\end{figure}

\begin{figure}
\resizebox{1\linewidth}{!}{\includegraphics*{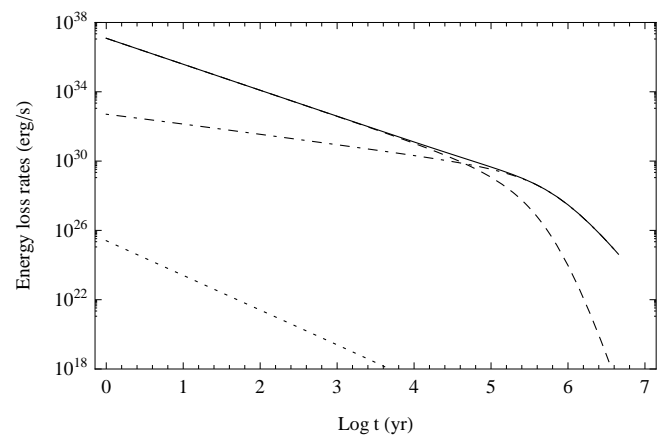}} \caption{The
energy lost by different processes in the neutron star with the hadronic equation of state, where the solid curve is the total energy loss rate, the dashed curve is the energy loss due to direct Urca neutrino emission, the dotted curve is due to modified Urca neutrino emission, and the dot-dashed curve is due to photon radiation.}\label{hadronic-energy}
\end{figure}

\begin{figure}
\resizebox{1\linewidth}{!}{\includegraphics*{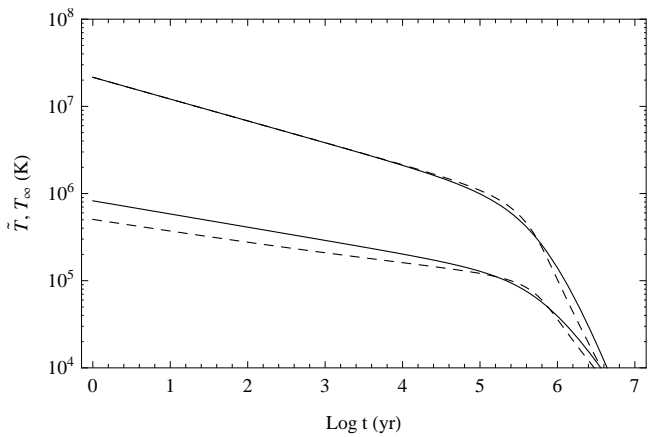}} \caption{The
temperature evolution of the neutron star with the hadronic equation of state with different $T_\infty$-$T$ relation. The solid curves correspond to $\eta=1$ while the dashed curves correspond to $\eta=10^{-20}$. The upper group is $\tilde{T}$ and the lower group is $T_\infty$.}\label{hadronic-eta}
\end{figure}

For the hadronic equation of state, Fig.~\ref{hadronic-T} shows
the evolution of the temperatures $\tilde{T}$ and $T_\infty$,
where it can be seen that the cooling process can be divided into
two stages, namely the neutrino-emission dominated and
photo-radiation dominated era. The two eras can be clearly seen
from the energy loss due to the different processes as shown in
Fig.~\ref{hadronic-energy}, where the switch from neutrino
emission to photon radiation occurs at
$t\approx3\times10^4\textrm{ yr}$. In Fig.~\ref{hadronic-energy}
the energy emission due to the modified Urca processes is also shown
demonstratively by considering only the processes involving the
nucleons in Eq.~(\ref{MUrca}). This, of course, underestimates the
energy loss due to the modified Urca processes, however, the
magnitude is expected to be of the same order. The energy loss by
these processes is seen to be less than $10^{-9}$ of the total
energy loss, such that the modified Urca processes are always
negligible in the present calculation. In
Figs.~\ref{hadronic-T} and \ref{hadronic-energy} the $T_\infty$-$T$
relation is used with $\eta=10^{-10}$. It is found that different values of
$\eta$ only slightly change the cooling process, as shown in
Fig.~\ref{hadronic-eta}, where two groups of cooling curves with
extremely large and almost vanishing $\eta$ are compared. The
difference is that, for the neutrino-emission dominated era,
$T_\infty$ can differ by about a factor of $2$ but the inner
temperature is almost not influenced. With less massive elements
in the envelop, i.e., smaller $\eta$, the turning point into a
photon-radiation dominated era is a little sharper and the final
temperatures are a little lower. In any case, these effects are
not very significant and can hardly be distinguished with the
present accuracy of observational data. Without loss of
generality, the value $\eta=10^{-10}$ will be used in this
section. Here we should also point out that the calculation cannot be carried out after $\tilde{T}$ is smaller than $10^4\textrm{
K}$, because then the $T_\infty$-$T$ relation given in
Ref.~\cite{Potekhin:1997} is no longer valid.

\begin{figure}
\resizebox{1\linewidth}{!}{\includegraphics{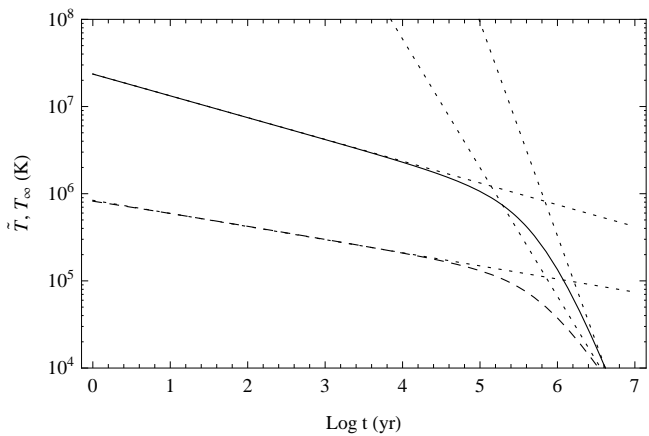}}
\caption{The same as in Fig.~\ref{hadronic-T} but with the mixed equation of state.}\label{hadronic-mix-T}
\end{figure}

\begin{figure}
\resizebox{1\linewidth}{!}{\includegraphics*{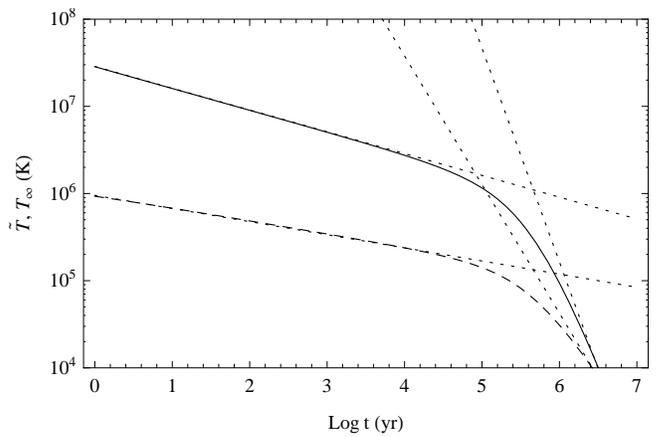}} \caption{The same as in Fig.~\ref{hadronic-mix-T} but with $B_c=170$MeV/fm$^3$.}\label{hadronic-mix170-T}
\end{figure}

The thermal evolution of the neutron star with the mixed equation of state is shown in Fig.~\ref{hadronic-mix-T}, which is
similar to the purely hadronic case. The straight lines represent
the asymptotic behavior, which are solved using Eq.~(\ref{therm})
by neglecting $Q_\gamma$ or $Q_\nu$ for the neutrino-emission
dominated or photon-radiation dominated era, respectively. For example, in the neutrino-emission dominated era, for the hadronic equation of state the temperatures scale with time as $\tilde{T}\sim1.6\times10^9\textrm{ }(t/{\rm
yr})^{-1/4}\textrm{K}$, $T_\infty\sim1.1\times10^7\textrm{
}(t/{\rm yr})^{-0.602/4}\textrm{K}$, where the power $0.602$
comes from the asymptotic approximation for the $T_\infty$-$T$
relation at low temperature, $T_\infty\propto T^{0.602}$. For the
mixed phase, the temperature scalings are just slightly higher:
$\tilde{T}\sim1.77\times10^9\textrm{ }(t/{\rm
yr})^{-1/4}\textrm{K}$, $T_\infty\sim1.13\times10^7\textrm{ }(t/{\rm
yr})^{-0.602/4}\textrm{K}$. This is due to the fact that the thermal
parameters, after being integrated over the whole star, are not
quite different for the two cases even though the stars have
different structure and mass. For example, for the hadronic case
we have $C_V=3.331\times10^{39}\tilde{T}_9\textrm{ erg/K}$ and
$Q_\nu=1.227\times10^{47}(\tilde{T}_9)^6\textrm{ erg/s}$, while
for the mixed phase $C_V=2.839\times10^{39}\tilde{T}_9\textrm{
erg/K}$ and $Q_\nu=7.280\times10^{46}(\tilde{T}_9)^6\textrm{
erg/s}$. To demonstrate the effect of the mixed phase, the same
curves are shown in Fig.~\ref{hadronic-mix170-T} with a smaller
bag constant of $B_c=170\textrm{ MeV}/\textrm{fm}^3$, where we see
that it causes a slightly higher temperature. As pointed out
earlier, a smaller bag constant causes the deconfined phase of
quarks to appear at lower densities, such that the volume of the
mixed phase is increased. In this case we find
$C_V=2.156\times10^{39}\tilde{T}_9\textrm{ erg/K}$ and
$Q_\nu=2.590\times10^{46}(\tilde{T}_9)^6\textrm{ erg/s}$.
Nevertheless, the total effect on the cooling process is not
drastically changed. According to these results, we expect that
even if the mixed phase is reduced because of screening and
surface effects, the cooling behavior is not changed considerably.

\begin{figure}
\resizebox{1\linewidth}{!}{\includegraphics{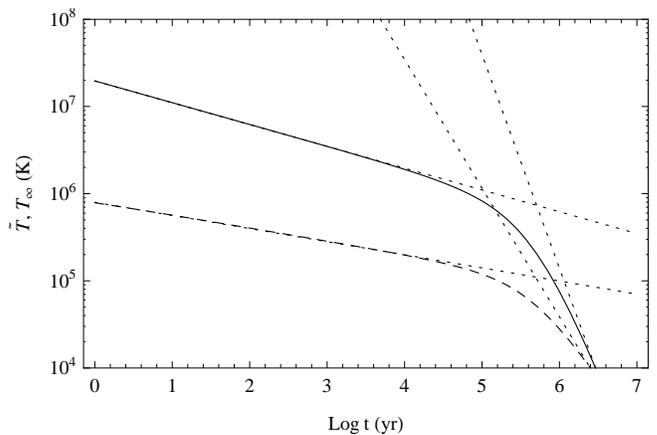}}
\caption{The same as in Fig.~\ref{hadronic-T} but with the nuclear equation of state.}\label{nuclear-T}
\end{figure}

\begin{figure}
\resizebox{1\linewidth}{!}{\includegraphics*{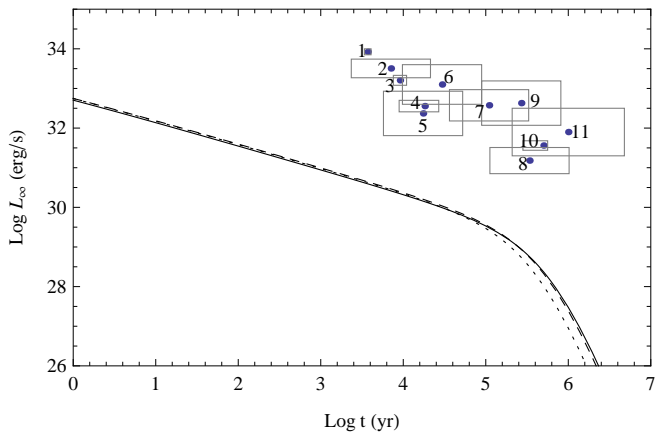}} \caption{The luminosity of neutron stars with different equations of state, with solid, dashed, and dotted curves representing the hadronic, the mixed, and the nuclear equation of state, respectively. The observational data are from Ref.~\cite{Page:2004}, with the numbers indicating the corresponding stars as: 1 - RX J0822-4247, 2 - 1E 1207.4-5209, 3 - RX J0002+6246, 4 - PSR 0833-45 (Vela), 5 - PSR 1706-44, 6 - PSR 0538+2817, 7 - PSR 0656+14, 8 - PSR 0633+1748 (Geminga), 9 - PSR 1055-52, 10 - RX J1856.5-3754, and 11 - RX J0720.4-3125.}\label{luminosity}
\end{figure}

For the nuclear equation of state, the maximum mass is
$1.719\textrm{ M}_{\odot}$ with $R=10.04\textrm{ km}$ and its
thermal evolution is shown in Fig.~\ref{nuclear-T}. To
summarize, the luminosity for the three types of equations of
state are plotted in Fig.~\ref{luminosity}, where a comparison
with observational data is also made. In all three cases the
cooling is too fast compared with the data. The star with the
nuclear equation of state cools a little faster, since the direct
Urca processes with hyperons in the hadronic phase and the mixed
phase are not as efficient as the nucleon direct Urca process
\cite{Prakash:1992}. However, the difference is quite small such
that it might be hard to distinguish these different equations of
state simply by their thermal evolution after the isothermal
condition is reached.

\section{Effect of superfluidity}

From the above results, we can see that our neutron stars cool too
fast because the direct Urca process is open for all the three
types of equations of state. An explanation could be that the
neutrino emissivity is overestimated by neglecting the possibility
of superfluidity. As pointed out in the introduction, it is
generally believed that superfluidity appears inside neutron stars
due to the attractive part of the nuclear force, for both the
nucleons and the hyperons. Due to the presence of an energy gap
$\Delta$, associated with the binding energy of the Cooper pairs,
the number of excitations near the Fermi surface are suppressed by
a factor of $\exp{(-\Delta/k_BT)}$ when the temperature is smaller
than a certain critical temperature $T_c$. Therefore, pairing
reduces the heat capacity and the neutrino emissivity
significantly when the star cools down to temperatures below $T_c$, thus the cooling behavior is drastically changed. In this section we will explore the effect of superfluidity to the cooling process of neutron stars.

\subsection{Superfluidity inside neutron stars}

In our present model for neutron stars, there are different fermions, namely the baryons, the leptons, and the quarks. The typical $T_c$ of baryons is about $0.1-1\text{ MeV}$, which is close to the initial temperature of neutron stars, such that the superfluidity of baryons is very important to the cooling process. As for the leptons, such as electrons, they could become superfluid due to interactions via phonons, however, the typical $T_c$ is very small, namely in the order of several kelvin. Hence lepton superfluidity is expected to be unimportant to the neutron
star cooling process. For quarks in the mixed phase, the gaps can
be about $50-100\textrm{ MeV}$ \cite{Page:2006}, which is much
higher than the typical temperatures of the stars we are concerned
with. The quark contribution will thus be suppressed more strongly
than the baryonic contribution. Without superfluidity the quark
contributions to the heat capacity and the neutrino emissivity
were shown to be of the same order as the baryonic ones, such that
with superfluidity they are completely negligible at the same
stellar temperature. Throughout this section we thus only consider
the contributions to $c_V$ and $q_\nu$ of non-superfluid leptons
and superfluid baryons but neglect those of the superfluid quarks.

Superfluidity not only reduces the neutrino emissivity but also
opens a different channel of neutrino emission based on the
breaking and formation of Cooper pairs. At temperatures not far
below $T_c$ thermal fluctuations can cause pair break-up into single-particle
excitations, which subsequently reform into pairs. Neutrino
emission due to such processes is quite efficient at temperatures
slightly below $T_c$ and can even surpass the direct Urca process
in some cases \cite{Yakovlev:1999rev}. However, its contribution
also decreases dramatically when $T$ becomes small due to the
exponential reduction factor mentioned previously. According to a
similar argument as above, we only need to consider such neutrino-emission processes involving baryon pairs.

By taking into account superfluidity, the dominant neutrino-emission process is no longer simply determined by the density,
which is crucial for the threshold of the direct Urca process, but
also depends on the reduced temperature $\tau=T/T_c$. However, it
is still true that the modified Urca process can be neglected as
long as the direct Urca process is open, since the modified Urca
process involves more particles and is more strongly suppressed due to the
energy gaps of all superfluid participants. Thus we again do not
consider the modified Urca process since the direct Urca process is always open in our study. According to Ref.~\cite{Yakovlev:1999rev} electron-electron bremsstrahlung
becomes a dominant process when all baryons are strongly superfluid at low temperature. So for a correct estimate of the neutrino emission at low temperatures we also include lepton-bremsstrahlung processes. Based on the discussion above, in order to include the effect of superfluidity we should recalculate $c_V$ and $q_\nu$ for all baryons, neglect those of the quarks in the mixed phase, keep the lepton contributions unchanged, and also include the newly introduced neutrino emission due to baryon pair break-up and lepton bremsstrahlung. We next discuss these effects separately.

To describe the superfluidity of the particles, it is necessary to
specify the type of pairing. At low density, the singlet-state
nuclear interaction is attractive, while at higher densities it
becomes repulsive. It is believed, however, that the triplet-state
interaction still provides an attractive channel at high
densities, such that triplet-state pairing is expected in the core
of the neutron star. This transition happens around the nuclear saturation
density $n_0=0.16\textrm{ fm}^{-3}$. The proton pairing is usually
taken to be in the singlet channel, even in the stellar core, due
to their low concentration. However, in our model, as seen from
Figs.~\ref{hadronicrn} and \ref{hadronicmixrn}, the densities of
neutrons and protons in the core are quite close to each other,
namely $n_p\sim0.2-0.3\textrm{ fm}^{-3}$ and $n_n\sim0.4\textrm{
fm}^{-3}$. It thus seems that both the protons and the neutrons
should form triplet-state pairs. The hyperons, which can also
become superfluid, are usually taken to pair in the singlet
channel. But in our model there are some hyperons, e.g.
$\Lambda^0$, which can have densities comparable to the neutrons
and protons. As was discussed earlier, the details of the
interactions between the hyperons are not well established and
thus there are some ambiguities in dealing with hyperons with
largely varying densities. For the singlet-state pairing of
neutrons, protons \cite{Page:2004}, and $\Lambda^0$'s
\cite{Balberg:1998} the results of many models largely
agree on the order of magnitude of the pairing gaps and their
density dependence. As discussed, triplet-state pairing is of
great interest to our model, however, there is little known about
this kind of pairing except for the case of neutrons for which it
is still highly model dependent. In fact, the existence of
triplet-state pairing inside neutron stars is still uncertain.
According to observations of the cooling neutron stars, it seems
necessary to have \textit{all} baryons in the superfluid state.
For example, the effect of superfluidity on the thermal evolution
of neutron stars with hyperons in the core has been studied by
Schaab {\it et al.} \cite{Schaab:1998}, which only included the
singlet-state pairing of $\Lambda^0$ at low densities but did not
include the triplet-state pairing of $\Lambda^0$ at high densities and
ignored the pairing of other hyperons. The cooling was found to be
too fast for heavy stars since not all the direct Urca processes
were suppressed.

To estimate the effect of superfluidity, we make the simplifying assumption that, regardless of the density, the neutrons, protons, and $\Lambda^0$'s pair in
the triplet channel, the remaining hyperons pair in the singlet
channel, and the critical temperatures of all the baryons are
taken to be equal. Equating all critical temperatures of the
various  baryons is consistent with our model, since all baryons
couple equally to the meson fields. Although our pairing mechanism
seems quite crude, we expect that, at least qualitatively, the
effect of pairing on the thermal evolution of the star will be
taken into account. To discuss our simplifications, we first need
to go into the details of how the specific heat and neutrino
emissivity are reduced by superfluidity.

\subsubsection{Reduction factor}

\begin{figure}
\resizebox{1\linewidth}{!}{\includegraphics{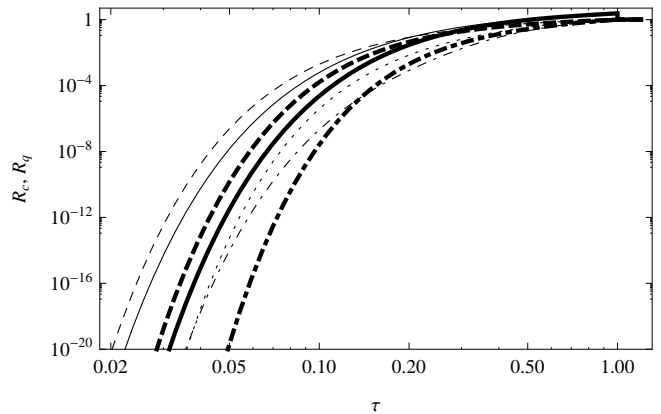}}
\caption{The reduction factors $R_c$ (solid curves) and $R_q$ (dashed curves) for different types of superfluidity, where type-A (thick curves) is always lower than type-B (thin curves) pairing. Notice the jump of $R_c$ at $\tau=1$. The dotted curve represents $R_{qB}^2$ used for the direct Urca process where both baryons are superfluid and have the same $T_c$. Also $R_{qAA}$ \cite{Levenfish:dual} (dot-dashed curve) and $R_{qA}^2$ (thick dot-dashed curve) are shown for comparison.}\label{reduction}
\end{figure}

The new expressions for $c_V$ or $q_\nu$ are easily obtained by
multiplying the original expressions by a reduction factor $R_c$
or $R_q$. These reduction factors are functions of the reduced
temperature $\tau$ and also depend on the type of superfluidity
considered. A rather systematic calculation of these reduction
factors has already been carried out (see review
\cite{Yakovlev:1999rev} and references therein). In general, the
different types of triplet-state pairing can be represented by the
projections  $m_J=0,\pm1,\pm2$ of the total angular momentum. We
only present the most studied $m_J=0$ case, which is denoted as
type-B pairing. Singlet-state pairing is referred to as type A. By
introducing the dimensionless variables $\tau=T/T_c$ and
$v(\tau)=\Delta(T)/k_BT$, the properties of superfluidity can be
described independent of $T_c$. Numerical fits for the energy gap
are given by
\begin{align}
v_A&=\sqrt{1-\tau}\left(1.456-\frac{0.157}{\sqrt{\tau}}+\frac{1.764}{\tau}\right),\\
v_B&=\sqrt{1-\tau}\left(0.7893+\frac{1.188}{\tau}\right),
\end{align}
and $v\equiv0$ at $\tau>1$. The reduction factors for $c_V$ are
\begin{align}
R_{cA}&=\left(0.4186+\sqrt{1.014+0.2510v_A^2}\right)^{2.5}\nonumber\\
&\times e^{1.456-\sqrt{2.120+v_A^2}},\\
R_{cB}&=\left(0.6893+\sqrt{0.6241+0.07975v_B^2}\right)^2\nonumber\\
&\times e^{1.934-\sqrt{3.740+v_B^2}}.
\end{align}
Notice that $R_{cA}=2.426$ and $R_{cB}=2.188$ are greater than $1$
at $\tau=1$, such that the specific heat is discontinuous as the
temperature falls below $T_c$. The neutrino emissivity for the
direct Urca process is different if both or only one of the two
involved baryons is superfluid. For the case with only one
superfluid baryon
\begin{align}
R_{qA}&=\left(0.2312+\sqrt{0.5911+0.02068v_A^2}\right)^{5.5}\nonumber\\
&\times e^{3.427-\sqrt{11.74+v_A^2}},\\
R_{qB}&=\left(0.2546+\sqrt{0.5556+0.01649v_B^2}\right)^5\nonumber\\
&\times e^{2.701-\sqrt{7.295+v_B^2}}.
\end{align}
Generally, type-A pairing always has a stronger reduction effect
than type B, as shown in Fig.~\ref{reduction}.

The reduction factor for the case where both particles are superfluid is not simply the product of the two reduction factors of each particle, due to different phase-space restrictions. This was pointed out in Ref.~\cite{Levenfish:dual}, where the authors presented the numerical fits for the reduction factors for the AA and BA cases. Here the cases are distinguished by the type of pairing (A or B) of the neutrons and protons, labeled as AA, BA or BB, where the first letter signifies the type of pairing for the neutrons and the second the pairing type for the protons. We do not present their complicated expressions here. The most important factor we need is for the BB case, which as far as we know has not been explicitly calculated yet. Therefore, we use $R_{qB}(\tau_1)R_{qB}(\tau_2)$ instead of $R_{qBB}(\tau_1,\tau_2)$ for those direct Urca processes with both baryons paired in the triplet-state, which we refer to as the standard setting. Furthermore, we even use this expression for \textit{all} the direct Urca processes with two superfluid baryons, independent of the pairing type. Therefore, the effect of superfluidity on the emissivity for the BB-type process is expected to be overestimated \cite{Yakovlev:1999rev}. On the other hand, since the above substitution is also applied to the $n$, $p$ and $\Lambda^0$ at lower densities and other hyperons, which are supposed to pair in the singlet channel, the suppression will be underestimated in these cases. The errors in these two approximations thus work in opposite direction and roughly reduce the total error of our calculation. The reduction factors of the direct Urca process with two superfluid baryons are compared in Fig.~\ref{reduction}. It is seen that their differences can become very large for $T\ll T_c$. However, in this case the reduction factors are extremely small and the lepton-bremsstrahlung processes are expected to be dominant, such that in this regime any ambiguity due to the type of pairing is expected to be unimportant.

\subsubsection{New neutrino emission processes due to superfluidity}

The neutrino emissivity associated with Cooper pair break-up is given by
\cite{Yakovlev:1999}
\begin{equation}\label{pairemi}
q^C_{\nu b}=1.17\times10^{21}N_\nu\frac{m_b^*\hbar k_b}{m_n^2c}a_bF(v)T_9^7 \textrm{ erg}/\textrm{cm}^{3}\textrm{s},
\end{equation}
where $N_\nu=3$ is the number of neutrino flavors, $a_b$ is a
numerical factor from the electroweak neutral currents and depends
on the quark composition of the baryon and the pairing type. As
far as we know, no calculation has been carried out which included
the triplet-state pairing of $\Lambda^0$. We simply use
$a_\Lambda=a_n$, considering that the quark composition of
$\Lambda^0$ is similar to the neutron with one $d$ quark replaced
by one $s$ quark, while their contributions to the neutral current
are the same. This treatment is different from the approximation
in Ref.~\cite{Okun:1984}, where the contribution from quarks other
than $u$ and $d$ is neglected. This may cause some
uncertainties, but what matters in the cooling process is the order
of magnitude of each process. Since $a_b$ appears as a multiplier
rather than an exponent in Eq.~(\ref{pairemi}), such an inaccuracy
will not be magnified during the calculation. In fact, we will see
that due to the assumption of a uniform $T_c$, the neutrino
emission associated with Cooper pair break-up is always negligible
compared to the direct Urca process even if the suppression due to
superfluidity is included. The factor $a_b$ for the various
baryons used in the calculation is shown in Table~\ref{paira}.
\begin{table}
\caption{The factor $a_b$ of the neutrino emissivity due to the Cooper pair break-up of the various particles \cite{Yakovlev:1999}. Notice that we take the coefficient of $\Lambda^0$ to be the same as $n$, which differs from Refs.~\cite{Yakovlev:1999,Okun:1984}.}
\begin{tabular}{|c|c|c|c|c|c|c|c|}\hline
&\multicolumn{3}{|c}{triplet-state}&\multicolumn{4}{|c|}{singlet-state}\\\hline
baryon&$p$&$n$&$\Lambda^0$&$\Sigma^\pm$&$\Sigma^0$&$\Xi^0$&$\Xi^-$\\\hline
$a_b$&$3.18$&$4.17$&$4.17$&$1.17$&$0$&$1$&$0.0064$\\\hline
\end{tabular}\label{paira}
\end{table}

The function $F(v)$ plays the same role as the reduction factor
and is fitted in Ref.~\cite{Yakovlev:1999} to:
\begin{align}
F_A=&(0.602v_A^2+0.5942v_A^4+0.288v_A^6)\bigg(0.5547+\nonumber\\
&\left.\sqrt{0.1983+0.0113v_A^2}\right)^{1/2}e^{2.245-\sqrt{5.04+4v_A^2}},\\
F_B=&\frac{1.204v_B^2+3.733v_B^4+0.3191v_B^6}{1+0.3511v_B^2}\bigg(0.7591+\nonumber\\
&\left.\sqrt{0.05803+0.3145v_B^2}\right)^2e^{0.4616-\sqrt{0.2131+4v_B^2}}.
\end{align}

Furthermore, the neutrino emissivity due to the lepton-bremsstrahlung process $ee$ is given by \cite{Kaminker:1999}
\begin{equation}
q_\nu^{ee}=2.089\times10^{11}\frac{\hbar ck_e}{1{\rm MeV}y_s}T_9^8 \textrm{ erg}/\textrm{cm}^{3}\textrm{s},
\end{equation}
where $y_s$ is a dimensionless parameter representing the effect
of screening in the plasma, $y_s=k_{sc}/2k_e$, with $k_{sc}$ the
screening wavenumber. This wavenumber is obtained for the case of
static screening in the limit of zero temperature \cite{Gnedin:1995} by the Thomas-Fermi expression:
\begin{equation}
k_{sc}^2=\frac{4e^2}{\pi\hbar^2}(\sum_lm^*_lk_l+\sum_bm^*_bk_bZ_b),
\end{equation}
where the summation over $b$ is over all charged baryons, and
$Z_b$ represents the effect of superfluidity on the baryons. Since
lepton bremsstrahlung only becomes important if all the baryons
are highly superfluid, for which $Z_b$ becomes negligibly small
compared to the lepton terms, we can simply omit the terms related
to the baryons. Subsequently, $y_s$ is simplified to
\begin{equation}
y_s=\sqrt{\frac{\alpha\mu_Q\sum_lk_l}{\pi\hbar ck_e^2}},
\end{equation}
where $\alpha$ is the fine-structure constant. Other processes
involving non-relativistic muons are calculated in
Ref.~\cite{Kaminker:1999}. In our model the muons are
relativistic, so we adopt the similarity criterion
\cite{Yakovlev:2001} and get
\begin{equation}
q_\nu^{e\mu}=4\frac{k_\mu}{k_e}q_\nu^{ee};\quad q_\nu^{\mu\mu}=\frac{k_\mu^2}{k_e^2}q_\nu^{ee}.
\end{equation}

\begin{figure}
\resizebox{1\linewidth}{!}{\includegraphics{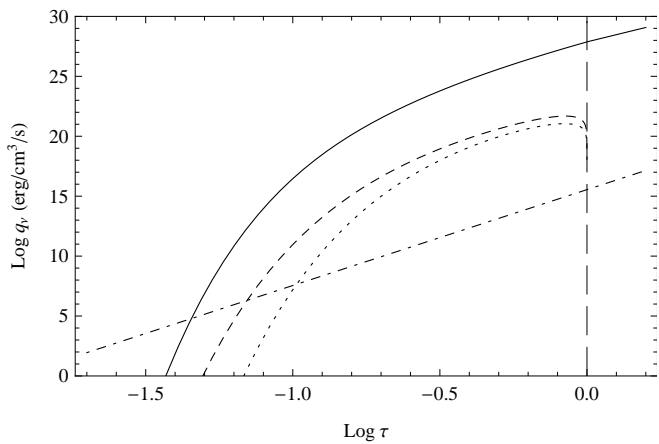}}
\caption{The neutrino emissivities of the various processes at different temperatures. The solid curve is for the direct Urca process, the dashed curve for the triplet-pair break-up process, the dotted curve is for the singlet-pair break-up process, while the dot-dashed curve for the lepton-bremsstrahlung process. The vertical dashed line at $\tau=1$ indicates the onset of superfluidity of all baryons. The emissivities are calculated at the central density of the maximum mass neutron star with the hadronic equation of state and $T_c=10^9\textrm{ K}$.}\label{qnupair}
\end{figure}

To summarize, in Fig.~\ref{qnupair} we present the neutrino
emissivities of the discussed processes as a function of reduced
temperature $\tau$ after the inclusion of superfluidity. Due to
our assumption of equal $T_c$, the emissivity associated with Cooper pair break-up is
always about $5$ orders of magnitude smaller than for the direct
Urca process. This agrees qualitatively with the result in
Ref.~\cite{Yakovlev:1999rev} for the equal $T_c$ case of nuclear
matter. For the neutron star with the mixed equation of state, the
situation is similar since the leading contribution of
triplet-state paired baryons changes very little. In fact, we can
always neglect the contribution to the neutrino emissivity
associated with Cooper pair break-up.

Finally, it should be pointed out that superfluidity can also
affect the equation of state. However, the equation of state used
in the TOV equations is an integral over all states in momentum
space and it is expected that the effect of the energy gap at the
Fermi surface is very small for very high densities. Therefore, we
can still use the stellar structure obtained from the previous
unaffected equation of state. Again, we only present the results
from the maximum mass star for each type of equation of state.

\subsection{Numerical results}

The calculation of the thermal evolution of the star is similar to
the case without superfluidity, but now the coefficients $C_V$ and
$Q_\nu$ in Eqs.~(\ref{CV}) and (\ref{Qnv}) include the
temperature-dependent reduction factors. Note that the relevant
parameter for the thermal evolution is $\tilde{T}$ rather than $T$, such that the reduction factors are functions of $\tau(r)=\tilde{T}e^{-\Phi(r)}/T_c$, which
is not constant throughout the star if we use a constant $T_c$.
This radial dependence should be included in the integration of
Eqs.~(\ref{CV}) and (\ref{Qnv}). Since the $\tau$ dependence of
the reduction factor is nonlinear, the variables $\tilde{T}$ and
$r$ cannot be separated, which means that the numerical integration
must be carried out at each instance to solve the
differential equation, cf.~Eq.~(\ref{therm}). To circumvent this we introduce an additional approximation by taking $T_ce^{\Phi(r)}$ a constant, such that
the reduced temperature $\tau$ is constant inside the star. This
decouples $\tau$ and $r$ in Eqs.~(\ref{CV}) and (\ref{Qnv}), which
means we do the radial integration only once. Then
Eq.~(\ref{therm}) is independent of the radial coordinate and the
reduction effect is represented by several extra $\tau$-dependent
terms in the coefficients. As is well known, the critical
temperature $T_c$ is a function of the density, however, the
dependence is often uncertain. Because of the discrepancies
between the pairing models, it is even unclear whether $T_c$ for
each type of baryon increases or decreases with baryon density in
the range of interest. In fact, setting $T_c=\textrm{const}$
throughout the star also implies some kind of density dependence
because the baryon density changes with the radius. In other
words, as an ansatz, $T_ce^{\Phi(r)}=\textrm{const}$ is as reasonable as $T_c=\textrm{const}$, even though the former seems
unnatural because of its dependence on the macroscopic stellar
structure. We use the former ansatz first because of
its simplicity. Later we make a comparison between these two
ans\"{a}tze. In any case, $e^{-\Phi(r)}$ only varies smoothly and
monotonically around $2$, for example, from about $2.2$ at $r=0$ to about $1.4$ at $r=R$ for the maximum mass neutron star with the hadronic equation of state,
so we expect the difference between these two ans\"{a}tze will not
be of several orders in magnitude.

In the numerical calculation with the first ansatz, we set
$T_ce^{\Phi(r)}=\tilde{T}(0)=10^9\textrm{ K}$ such that the
baryons are superfluid from the start. This implies we have
$T_c\approx2.2\times10^9\textrm{ K}$ at the central density while
$T_c\approx1.4\times10^9\textrm{ K}$ near the surface. If,
however, $\tilde{T}(0)$ were to be set higher, the reduction
factors would at first not take effect and the direct Urca process
would thus quickly reduce $\tilde{T}$ down to $T_c$ after which it
would be suppressed due to the onset of superfluidity and the
cooling slowed down. This process takes only several seconds. On
the other hand, if $\tilde{T}(0)$ would have been slightly lower,
for example, above $0.1T_c$, then in the beginning the superfluid
suppression would not be very strong and the direct Urca process
could still cool the star to low temperature, i.e., the highly
superfluid regime, within one year. Since the isothermal
assumption we used here is virtually a later stage behavior, given
a reasonable $T_c$ and the well-accepted range of $\tilde{T}(0)$
of neutron stars, the result we presented in this section depends
little on the accurate values of the temperature. As
before, we take the neutron star with the hadronic equation of
state as the basic example, then we study the stars with the mixed
and nuclear equation of state.

\begin{figure}
\resizebox{1\linewidth}{!}{\includegraphics{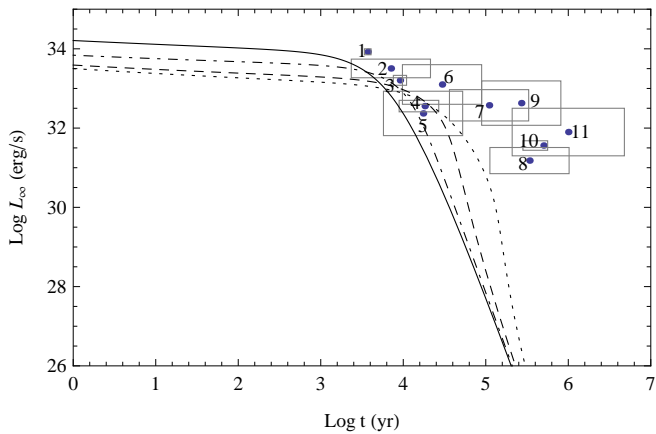}}
\caption{The luminosity of a neutron star with the hadronic equation of state. The solid, dot-dashed, dashed, and dotted curves are for $\eta=10^{-9},10^{-11},10^{-13}$, and $10^{-17}$, respectively. The observational data are again shown for comparison.}\label{etacompare}
\end{figure}

We find several general properties of the cooling behavior of
stars containing superfluid matter from the numerical results. One
obvious difference with the non-superfluid case is the shape of
the cooling curve. Since the neutrino-emission processes, which
dominate the early stage cooling, are strongly suppressed with
decreasing temperature, the temperature stays high for a longer
time and the shift to the photon radiation era takes place at
higher temperatures. In contrast to the large slope in the
non-superfluid case, we find a rather slowly decreasing plateau
around $\tilde{T}\sim10^8\textrm{ K}$ and
$T_\infty\sim10^6\textrm{ K}$, followed by a much sharper turn to
the photon radiation stage as can be seen in the following
figures. Another difference can be seen in that the parameter
$\eta$ in the $T_\infty$-$T$ relation plays a more significant
role. Because the neutrino-emission processes are suppressed due
to superfluidity, photon radiation becomes more important at
higher temperatures and the dependence on how the inner
temperature is screened by the surface layer becomes more visible.
We present the luminosity curves with different $\eta$ in
Fig.~\ref{etacompare}, where we see that the observational data
favor moderate values of $\eta$. In the following, we usually
use $\eta=10^{-13}$.

\begin{figure}
\resizebox{1\linewidth}{!}{\includegraphics{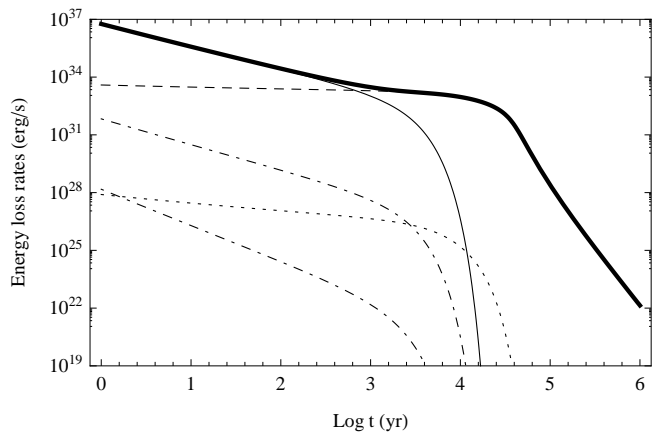}}
\caption{The energy lost by the various processes in the neutron star with the hadronic equation of state, where the thick solid curve is the total energy loss rate, the thin solid curve is for the direct Urca process, the two dot-dashed curves are for the Cooper pair break-up processes with the upper one for the triplet-state pairing and the lower one for the singlet-state pairing, the dotted curve is for lepton bremsstrahlung, and the dashed curve is for photon radiation.}\label{hadronic-energy-sf}
\end{figure}

\begin{figure}
\resizebox{1\linewidth}{!}{\includegraphics{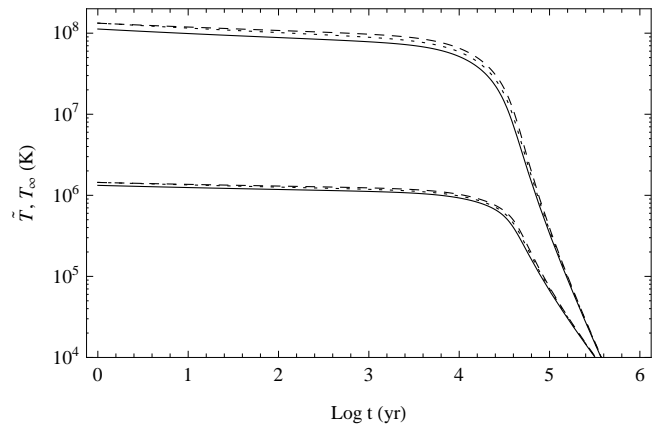}}
\caption{The temperature evolution of the neutron star with the hadronic equation of state. The various curves represent different pairing types. The solid curves are for our standard setting, the dotted curves are when all baryons pair in the singlet-state, and the dashed curves are also for singlet-state pairing but with the reduction factor $R_{AA}$ replaced by $R_A^2$. As before, the upper group is $\tilde{T}$ and the lower group is $T_\infty$, respectively.}\label{paircompare}
\end{figure}

Here some other general properties are summarized. First, we find that
the heat capacity $c_V$ of the baryons is quickly reduced as a
consequence of superfluidity. Starting from $t\approx1\textrm{
yr}$, the heat capacity is completely dominated by the lepton
contribution, such that the pairing type of the baryons is seen to
be of no importance to the heat capacity. Second, as was already
shown in Fig.~\ref{qnupair}, neutrino emission via Cooper pair break-up
is always negligible compared to the direct Urca process. We also
find that the role of lepton bremsstrahlung in the cooling process
is negligible since in the later stages when it dominates the
neutrino emission, photon radiation has already become the
dominant cooling effect. Therefore, the cooling process is
determined only by the direct Urca process and photon radiation.
The various energy loss rates are shown in
Fig.~\ref{hadronic-energy-sf}. Third, the direct Urca process
depends on the type of pairing. However, this dependence is quite
moderate. We show the different cooling curves for the various
pairing types in Fig.~\ref{paircompare}. For singlet-state pairing, we
see that neutrino emission is more strongly suppressed and the
resulting temperature is slightly higher than the other types of
pairing. This result confirms that the cooling behavior is not
strongly dependent on the type of pairing of the baryons.

By comparing Fig.~\ref{etacompare} and Fig.~\ref{luminosity}, we
see that the luminosity curves in the superfluid case are much
closer to observations than in the non-superfluid case. However,
the curves do not agree with all the observational data except for
the younger stars, namely number 1-5. This mismatch may lie in some
important mechanisms which are neglected in the present
calculation. For example, the magnetic field of the neutron star
can affect the photon radiation at the surface \cite{Page:2006}
such that it no longer obeys the standard black-body radiation law
and shift the photon-radiation era to a later time
\cite{Page:1996}. This reduction of photon radiation also decreases the
luminosity, which is unfavorable according to the cooling
curves shown in Fig.~\ref{etacompare}. However, there are some
known heating mechanisms which are not considered here, for
example, due to magnetic field energy and rotational energy (for
more details, see Ref.~\cite{Page:2006} and references therein),
such that the temperature and the luminosity can remain higher. Besides, note that sources such as number 10 and 11 may be old magnetars. As a consequence, the age estimates may not be correct, and the cooling history may be anomalous. Both are a result of the decay of the strong magnetic field: age estimates assume magnetic braking of rotation with a constant magnetic field, whereas the decay of the magnetic field results in heating of the neutron star \cite{Heyl:1998}. We discuss these issues in future work.

\begin{figure}
\resizebox{1\linewidth}{!}{\includegraphics{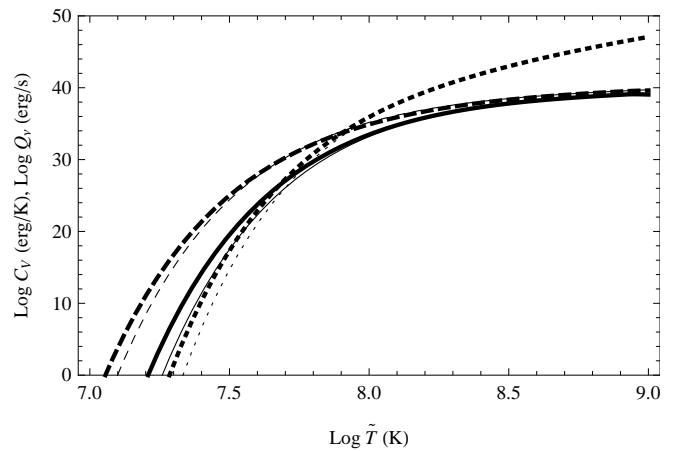}}
\caption{The temperature dependence of $C_V$ (solid curves for the singlet-state pairing contribution and dashed curves for the triplet-state pairing contribution) and $Q_\nu$ (dotted curves) calculated with the two different ans\"{a}tze for $T_c$ in the neutron star with the hadronic equation of state. For each quantity, the thick higher curve is for $T_c=\textrm{const}$ and the thin lower curve is for $T_ce^{\Phi(r)}=\textrm{const}$, respectively.}\label{ansatzcompare}
\end{figure}

\begin{figure}
\resizebox{1\linewidth}{!}{\includegraphics{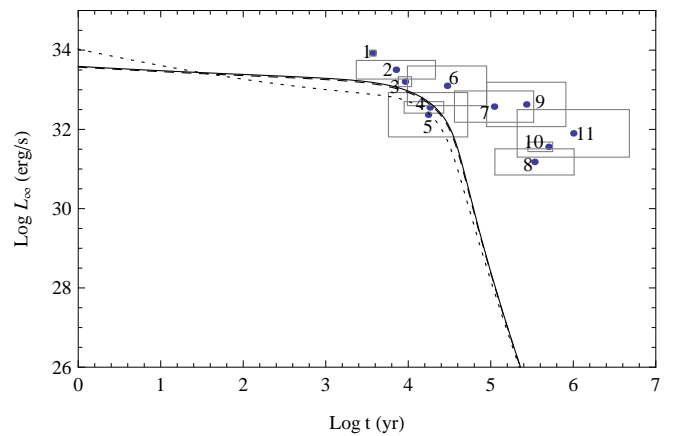}}
\caption{The luminosity calculated with two different ans\"{a}tze for $T_c$ in the neutron star with the hadronic equation of state. The solid curve is for $T_ce^{\Phi(r)}=10^9\textrm{ K}$ and the dashed curve is for $T_c=1.9\times10^9\textrm{ K}$, respectively. For comparison, the result with the ansatz $T_c=10^9\textrm{ K}$ is shown by the dotted curve. The observational data are again shown.}\label{ansatzluminositycompare}
\end{figure}

Now we discuss the second ansatz, $T_c=\textrm{const}$. Since
$e^{-\Phi(r)}>1$, if we set $T_c=\tilde{T}(0)=10^9\textrm{ K}$ we
will have a lower $T_c$ than in the previous ansatz and the
baryons will not be superfluid everywhere inside the star at
$t=0$. To make a proper comparison between the two, we set $T_c=1.9\times10^9\textrm{ K}$, where the extra factor
$1.9$ is the middle value of $e^{-\Phi(r)}$ inside the star (the
outer value is taken at $r=7\textrm{ km}$ rather than at the surface
since the direct Urca process starts around here, see
Figs.~\ref{hadronictherm} and \ref{hadronicmixtherm}). With this
value of $T_c$, at $t=0$, the baryons near the surface are superfluid
while those in the core are not, since
$\tau=\tilde{T}e^{-\Phi(r)}/T_c$ is bigger in the core.
Nevertheless we expect the integrated results will be comparable
to those of the previous ansatz. As a comparison, in
Fig.~\ref{ansatzcompare}, $C_V$ and $Q_\nu$ are shown for these
two ans\"{a}tze. We see that the difference between the two
increases with decreasing temperature and only becomes significant
when their values are very small. Therefore, it is not surprising
to find that their cooling curves are similar, as is shown in
Fig.~\ref{ansatzluminositycompare}. The two curves are quite close
except that the second ansatz has a little lower luminosity, because with
$T_c=\textrm{const}$ the higher temperature in the core delays
the suppression of the direct Urca process, such that the total
cooling rate is larger in the beginning. Furthermore, the result of the ansatz $T_c=10^9\textrm{ K}$ mentioned above is also shown, where we can see that the high cooling rate due to the direct Urca process remains for a longer time and the luminosity is even more reduced after the neutron star has become completely superfluid. Here we see again that different initial conditions have little influence on the later-stage behavior.

\begin{figure}
\resizebox{1\linewidth}{!}{\includegraphics{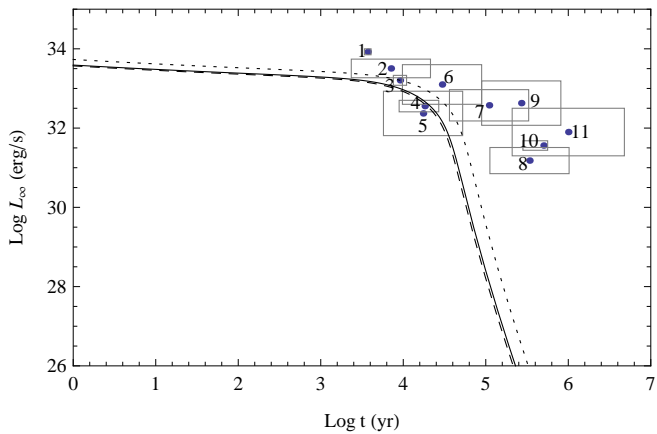}}
\caption{The luminosity of the neutron stars with the various equations of state with the effect of superfluidity included. The solid, dashed, and dotted curves represent the hadronic, mixed, and nuclear equation of state, respectively. The observational data are again presented.}\label{luminositysf}
\end{figure}

Finally, we briefly present the results for the neutron stars
with the mixed and nuclear equation of state. For the nuclear
case, we always take the neutrons and protons to pair in the
triplet channel. The general arguments still apply, i.e., we only
need to consider the direct Urca process and the photon radiation
as the dominant cooling mechanisms. Without too much difference,
the simple ansatz $T_ce^{\Phi(r)}=\textrm{const}$ will be used.
The cooling behavior is mainly determined by the heat capacity,
the direct Urca neutrino emissivity, and the photon radiation. These
three parameters are not quite different among the various
equations of state, even including the effect of superfluidity. The dominant contribution to the heat capacity is
due to the leptons. The direct Urca process is dominated by the
triplet-state paired baryons which make up the majority of the
star and are less affected by superfluidity. The photon radiation
is again just black-body radiation. Therefore, we expect that the
cooling process should be similar in these three cases, whose
luminosity curves are shown in Fig.~\ref{luminositysf}. Compared
to Fig.~\ref{luminosity}, besides the overall increase in the
luminosity as previously discussed, there is some difference in
the order of magnitude of the luminosity among these three equations of
state. With superfluidity, the nuclear case always has the largest luminosity, which is due to the higher lepton fraction in the star. We
find that the heat capacity of the leptons in the neutron star with the
nuclear equation of state is about twice as large as that for the
star with the hadronic or mixed equation of state. This larger lepton heat capacity becomes significant when the baryons (and the quarks) are highly superfluid. As can be seen in Eq.~(\ref{therm}), the larger the heat capacity is, the higher the corresponding temperature should be. This explains the largest luminosity of the nuclear case in Fig.~\ref{luminositysf}. Another difference is that, with superfluidity, the luminosity of the mixed case is always smaller than that of the hadronic case, while for the non-superfluid case as in Fig.~\ref{luminosity}, there is no such clear order. The reason also lies in the effect of superfluidity on the heat capacity, since in the mixed case the quark contribution is strongly suppressed due to superfluidity and completely neglected during our calculation. Of course, the neutrino emissivity from quarks is also reduced, but the reduction in the heat capacity is relatively stronger. This is because, without superfluidity, quarks contribute less to the neutrino emissivity than baryons, but their contribution to the heat capacity is comparable.

\section{Conclusion and discussion}

We have carefully compared the thermal evolution of different
types of neutron stars, namely with the hadronic, the mixed phase
of hadronic and strange quark matter, and the nuclear
equation of state. We find that the direct Urca process is open in
all of these cases and thus results in relatively fast cooling
behavior. Although the details concerning the heat capacity and
neutrino emissivity can be rather different in these cases, the
cooling curves are quite similar after the stars become
isothermal. However, the behavior in the early stages before the
stars become isothermal could be significantly different, but such a study
requires the knowledge of the thermal conductivity of these
complex systems. The geometrical structure of the mixed phase is
also expected to play an important role, although no decisive
conclusion has yet been drawn.

As we have seen, the fast cooling in the non-superfluid case did not
agree with observations. In order to remedy this discrepancy,
superfluidity was introduced, which significantly reduces the
efficiency of the direct Urca process as well as the heat
capacity. The resulting cooling curve is much closer to the
observational data. We also found that the particular pairing type
of the superfluid baryons is not very important to the thermal
evolution of the star. The thermal evolution after the star
becomes isothermal is not strongly dependent on the initial
temperature of the star or on the critical temperature related to
superfluidity, as long as they are within reasonable ranges. The
robustness of the results with superfluidity is quite helpful in
order to remove the uncertainties concerning baryon
superfluidity at high density. Note that the cooling process is
almost completely determined by the direct Urca process and photon
radiation even after including the effects due to
superfluidity. The cooling curves of the neutron stars with the
three equations of state are still quite similar when
superfluidity is included. We expect that a calculation including the
magnetic field and rotation gives an even better
agreement with the observational data. Besides, since our nuclear model is geared originally towards nuclear matter near the saturation point, a further improvement of it may play an important role in getting a better agreement with the observations. By incorporating the properties at higher densities, e.g., the coupling constants for hyperons, it is possible to get cooling curves covering most of the observational data, as in Ref. \cite{Page:2006}.

\section*{Acknowledgments}

We thank Erik Laenen for helpful discussions during the early stage of this work. This work is supported by the Stichting voor Fundamenteel Onderzoek der Materie (FOM) and the Nederlandse Organisatie voor Wetenschaplijk Onderzoek (NWO).

\end{document}